\documentclass[manuscript]{aastex}
\usepackage{amsmath}
\usepackage{verbatim}
\usepackage{subfigure}
\usepackage{epsfig}
\usepackage{graphicx}
\newcommand{\vy}[2]{#1_{\scriptscriptstyle #2}}
\newcommand{\citen}[1]{\citeauthor{#1} \citeyear{#1}}

\def\gtorder{\mathrel{\raise.3ex\hbox{$>$}\mkern-14mu
             \lower0.6ex\hbox{$\sim$}}}
\def\ltorder{\mathrel{\raise.3ex\hbox{$<$}\mkern-14mu
             \lower0.6ex\hbox{$\sim$}}}
\def\proptwid{\mathrel{\raise.3ex\hbox{$\propto$}\mkern-14mu
             \lower0.6ex\hbox{$\sim$}}}
\def\0946{PG~0946+301}

\def\arcsec{\ifmmode '' \else $''$\fi}

\def\arcsecpoint{\ifmmode ''\!. \else $''\!.$\fi}

\def\kms{\ifmmode {\rm km\ s}^{-1} \else km s$^{-1}$\fi}
\def\Msun{\ifmmode {\rm M}_{\odot} \else M$_{\odot}$\fi}
\def\Lsun{\ifmmode {\rm L}_{\odot} \else L$_{\odot}$\fi}
\def\Zsun{\ifmmode {\rm Z}_{\odot} \else Z$_{\odot}$\fi}

\def\ergscm2{ergs\,s$^{-1}$\,cm$^{-2}$}
\def\icm3{{\rm cm}^{-3}}
\def\icm2{{\rm cm}^{-2}}
\def\qo{\ifmmode q_{\rm o} \else $q_{\rm o}$\fi}
\def\Ho{\ifmmode H_{\rm o} \else $H_{\rm o}$\fi}
\def\ho{\ifmmode h_{\rm o} \else $h_{\rm o}$\fi}
\def\ltsim{\raisebox{-.5ex}{$\;\stackrel{<}{\sim}\;$}}
\def\gtsim{\raisebox{-.5ex}{$\;\stackrel{>}{\sim}\;$}}
\def\vFWHM{\ifmmode v_{\mbox{\tiny FWHM}} \else
            $v_{\mbox{\tiny FWHM}}$\fi}
\def\CCF{\ifmmode F_{\it CCF} \else $F_{\it CCF}$\fi}
\def\ACF{\ifmmode F_{\it ACF} \else $F_{\it ACF}$\fi}
\def\Halpha{\ifmmode {\rm H}\alpha \else H$\alpha$\fi}
\def\Hbeta{\ifmmode {\rm H}\beta \else H$\beta$\fi}
\def\Hgamma{\ifmmode {\rm H}\gamma \else H$\gamma$\fi}
\def\Hdelta{\ifmmode {\rm H}\delta \else H$\delta$\fi}
\def\Lya{\ifmmode {\rm Ly}\alpha \else Ly$\alpha$\fi}
\def\Lyb{\ifmmode {\rm Ly}\beta \else Ly$\beta$\fi}
\def\Lyg{\ifmmode {\rm Ly}\beta \else Ly$\gamma$\fi}
\def\hi{H\,{\sc i}}

\def\ci{C\,{\sc i}}
\def\cii{C\,{\sc ii}}
\def\ciii{\ifmmode {\rm C}\,{\sc iii} \else C\,{\sc iii}\fi}
\def\civ{\ifmmode {\rm C}\,{\sc iv} \else C\,{\sc iv}\fi}

\def\niii{N\,{\sc iii}}
\def\niv{N\,{\sc iv}}
\def\nv{N\,{\sc v}}
\def\oi{O\,{\sc i}}
\def\oii{O\,{\sc ii}}
\def\oiii{O\,{\sc iii}}
\def\o5007{[O\,{\sc iii}]\,$\lambda5007$}
\def\oiv{O\,{\sc iv}}
\def\ov{O\,{\sc v}}
\def\ovi{O\,{\sc vi}}

\def\oviii{O\,{\sc viii}}

\def\nev{Ne\,{\sc v}}
\def\nevi{Ne\,{\sc vi}}
\def\neviii{Ne\,{\sc viii}}
\def\naix{Na\,{\sc ix}}

\def\mgx{Mg\,{\sc x}}
\def\siiv{Si\,{\sc iv}}

\def\siII{Si\,{\sc ii}}
\def\sixii{Si\,{\sc xii}}
\def\sixiv{Si\,{\sc xiv}}

\def\sii{S\,{\sc ii}}

\def\siv{S\,{\sc iv}}

\def\svi{S\,{\sc vi}}
\def\sxvi{S\,{\sc xvi}}

\def\feii{Fe\,{\sc ii}}

\def\alii{Al\,{\sc ii}}

\def\o{\o}
\begin{document}
%\vspace{-1in}

%\title{A NEW WINDOW ON AGN FEEDBACK, QUASAR OUTFLOWS IN THE FAR UV: 
%       HST/COS OBSERVATIONS OF QSO~HE0238--1904\altaffilmark{*}\\
\title{QUASAR OUTFLOWS AND AGN FEEDBACK IN THE FAR UV: 
       HST/COS OBSERVATIONS OF QSO~HE0238--1904\altaffilmark{*}}
%\today}

\author{
 Nahum Arav\altaffilmark{1}, 
 Benoit Borguet\altaffilmark{1},
 Carter Chamberlain\altaffilmark{1}, 
 Doug Edmonds\altaffilmark{1},
 Charles Danforth\altaffilmark{2},
}

\altaffiltext{*}{Based on observations made with the Hubble Space Telescope }
\altaffiltext{1}{Department of Physics, Virginia Tech, Blacksburg, VA 24061: arav@vt.edu}
 \altaffiltext{2}{CASA, University of Colorado, 389 UCB, Boulder, CO 80309-0389}

\begin{abstract}
Spectroscopic observations of quasar outflows at rest-frame 500\AA--1000\AA\  have immense diagnostic power.
We present analyses of such data, where  absorption
troughs from three important ions are measured:
first,  \oiv/\oiv* that allow us to obtain the distance of high ionization
outflows from the AGN;
second, \neviii\ and \mgx\ that are sensitive to the very high ionization phase of the outflow.
Their inferred column densities, combined with those of troughs from  \ovi,
\niv, and \hi, yield two important results: 1) The outflow shows two ionization
phases, where the high ionization phase carries the bulk of the material.
This is similar to the situation seen in x-ray warm absorber studies. Furthermore, 
the low ionization phase is inferred to have a volume filling factor of $10^{-5}-10^{-6}$.  2) From
the \oiv*/\oiv\ column density ratio, and the knowledge of the ionization
parameter, we determine a distance of 3000 pc\ from the outflow to the central
source.  Since this is a typical high ionization outflow, we can determine robust values 
for the mass flux and kinetic luminosity of the outflow: 40 \Msun\ yr$^{-1}$
and $10^{45}$ ergs s$^{-1}$, respectively, where the latter is roughly
equal to 1\% of the bolometric luminosity. 
Such a large kinetic luminosity and mass flow rate measured in a typical high
ionization wind suggests that quasar outflows are a major contributor to AGN
feedback mechanisms.

% quasar outflows based on high resolution
% spectroscopic 
% data shortwards of the rest-frame Lyman limit

\end{abstract}


\section{INTRODUCTION}

%%%% NEW %%%%

Quasar outflows are detected as UV absorption troughs blueshifted with
respect to the AGN's rest frame spectrum.  The ubiquity and wide opening angle deduced from the detection rate of
these mass outflows, allows for efficient interaction with the surrounding medium.
The energy, mass, and momentum carried by these outflows are thought to play an
important role in shaping the early universe and dictating its evolution
\citep[e.g.][]{Scannapieco04,Levine05,Hopkins06,Cattaneo09,Ciotti09,Ciotti10,Ostriker10}.
Theoretical studies and simulations show that this so-called AGN feedback can provide
an explanation for a variety of observations, from the chemical enrichment
of the intergalactic medium, to the self-regulation of the growth of the
supermassive black hole and of the galactic bulge \citep[e.g.][and references
therein]{Silk98,dimatteo05,Germain09,Hopkins09,Elvis06}.

In order to assess this contribution quantitatively, we initiated a research
program to determine the mass flow rate  ($\dot{M}$) and kinetic luminosity
($\dot{E}_k$) of the outflows.  We began by analyzing data from low-ionization outflows 
(showing \feii/\feii* and \siII/\siII* troughs: 
 \citep{Arav08,Korista08,Moe09,Dunn10a,Bautista10,Aoki11,Borguet12a}.  
While doing so we built upon ours \citep{deKool01, deKool02,deKool02b} and other groups \citep{Wampler95,Hamann01} previous 
investigations of such outflows.  
As we discuss below, the rarity of these low-ionization outflows put a significant uncertainty on the extrapolation of 
their results to the majority of quasar outflows, which show higher ionization species. Therefore, we recently started analyzing 
outflows that show troughs from the high ionization \siiv/\siiv* species \citep{Dunn12,Borguet12b}. Where we had to overcome the difficulty of 
these troughs being heavily contaminated by intervening \Lya\ forest absorption in spectra of high redshift quasars. These efforts 
yielded the most energetic BALQSO outflow measured to date, with a
kinetic luminosity of at least $10^{46}$ ergs s$^{-1}$, which is 5\% of the
bolometric luminosity of that high Eddington ratio quasar \citep{Borguet13}.

In this paper we explore another promising part of the spectrum for quasar outflow science, the rest-frame $500-1000$ \AA.
Earlier studies of quasar outflows in this spectral regime utilized some of the diagnostic power of this band
(e.g., Q 0226$-$1024, \citen{Korista92}; 
UM 675, \citen{Hamann95}; 
Q SBS 1542+541, \citen{Telfer98};  
PG 0946+301,  \citen{Arav99a}, \citen{Arav01};  
J2233$-$606, \citen{Petitjean99}; 
3C 288.1, \citet{Hamann00};
HE 0226$-$4110, \citen{Ganguly06}). 
However, the UV spectrographs used in these works lacked the combination of spectral resolution and sensitivity 
that allows the full diagnostic potential of outflow absorption troughs to be used (see Section 2.2).  
For example, none of these studies were able to measure the distance of the outflows from the central source, which is crucial for 
determining the relationship between the outflow and the host galaxy.

The Cosmic Origins Spectrograph (COS) onboard the Hubble Space Telescope (HST) has improved this situation dramatically. 
We can now obtain high enough quality data on bright, medium-redshift quasars ($1.5\gtorder z \gtorder 0.5$) 
to allow for detailed and reliable analysis of quasar-outflows spectra in this regime.
Pioneering work on such data was done by \citet[hereafter M2012]{Muzahid12}  who analyzed 
high resolution HST/COS observations (supplemented by FUSE archival data) of quasar HE0238--1904. Their main finding was
that the outflowing gas must have two ionization phases. 
Here we detail a careful reanalysis of the same data set, where we are able to:
a) detect absorption features associated with \oiv*, 
which allows us to determine the distance, $\dot{M}$ and $\dot{E}_k$ of the outflow and b) better quantify the two 
ionization phases of the outflow and put limits on its metalicity.

The plan of this paper is as follows:
In Section~2 we discuss the diagnostic power of troughs from
the rest wavelength between $500-1000$ \AA\ 
compared with those of longward wavelength troughs.
COS observations of HE0238--1904 and the absorption troughs' characterization are described in Section~3 and Section~4.
The troughs' column density measurements
are described in Section~5.
Photoionization modeling is discussed in Section~6, and the energetics of the outflow in Section 7. 
In Section 8 we elaborate on the two-phase nature of the outflow, 
determine its distance from the central source, and derive the 
$\dot{M}$ and $\dot{E}_k$ of the outflow.  We
discuss our results in Section~9 and summarize them in Section~10.

\section{DETERMINING THE PHYSICAL CONDITIONS AND ENERGETICS OF QUASAR OUTFLOWS}

\subsection{From Absorption Spectrum To Kinetic Luminosity Determination}

Assuming the outflow is in the form of a partial thin spherical shell ($\Delta
R/R\ltorder1/2$)
moving with velocity  $v$, it's mass ($M$), mass flow rate  ($\dot{M}$) and kinetic luminosity
($\dot{E}_k$) are given by (see discussion
in \citealt{Borguet12a}):
 \begin{equation}
M\simeq 4\pi \Omega R^2 \vy{N}{H}\mu m_p\ \ \ \ \
 \dot{M}\equiv \frac{M}{(R/v)}=4\pi \Omega R \vy{N}{H}\mu m_p v \ \ \ \ \ 
\dot{E}_k=\frac{1}{2}\dot{M} v^2,
\label{energetics} 
\end{equation}
where $\Omega$ is the fraction of the total solid angle
occupied by the outflow, $R$ is the distance of the outflow from the
central source, $\vy{N}{H}$ is the total hydrogen column density of
the outflow, $m_p$ is the mass of the proton and $\mu=1.4$ is the
molecular weight of the plasma per proton. We note that $\dot{M}$ is the average mass flow rate over the dynamical time-scale ($R/v$).

To measure the quantities given in equation (\ref{energetics}) the following steps are
needed: \\
 {\bf 1.} Measuring reliable ionic column densities ($N_{ion}$) from
the observed troughs (see Section~5). \\
 {\bf 2.} Photoionization modeling to determine $\vy{N}{H}$ and the ionization
parameter ($U_H$) from the measured $N_{ion}$
  (Section~6). \\
 {\bf 3.} Determining $R$: a) measuring the number
density ($\vy{n}{H}$) via troughs from metastable levels (Section~5) and b) using the
inferred
$\vy{n}{H}$ and $U_H$ to solve for the distance of the absorbing gas (Section~7). \\
 {\bf 4.} Constraining the solid angle $\Omega$ of the outflow (Section~7).  \\
 {\bf 5.} Measuring or constraining the chemical abundances (Section~6).  \\
The last variable $v$, is easily measured from the Doppler shift of the
absorption trough with respect to the systemic redshift of the quasar.

Over the past decade  significant advancements were made on all these steps 
{\bf 1.} \citep{Arav97,Arav99a,Arav99b,Arav02,Arav05,Arav08,Arav12,Hamann97b,deKool02,Gabel03,Gabel05a,Moe09},
{\bf 2.} \citep{Arav01,Arav01b,Arav07,Gabel05b,Korista08,Edmonds11}
{\bf 3.} \citep{deKool01,Hamann01,deKool02,deKool02b,Moe09,Dunn10a,Bautista10,Borguet12a,Borguet13}
{\bf 4.} \citep{Hewett03,Ganguly08,Dai08,Knigge08,Dunn12,Dai12},
and {\bf 5.} \citep{Arav01b,Arav07,Gabel06,Moe09,Borguet12b}.

However, due to astrophysical and instrumental constraints, most of these
investigations dealt with \Lya\ (1215\AA) and longer wavelength absorption
troughs, while  a small number of low luminosity and low redshift objects were
studied down to 1000\AA\ rest wavelength. Unfortunately, the rest wavelength
$\gtorder1000$ \AA\ is rather limited in spectral diagnostics for two important
aspects of the outflows. 

First, the highest ionization species available at these wavelength is \ovi. This
does not allow us to probe the higher ionization material that is known to exist
in these outflows via x-ray observations.  For example, the Chandra/FUSE/HST-STIS
campaign on Mrk 279 found clear evidence for two ionization components in the
outflow separated by roughly a factor of a hundred in ionization parameter. Only
the low ionization 
phase was detected in the UV observations (which covered \ovi\ absorption),
where the independent UV and x-ray analyses were in very good agreement for the
$\vy{N}{H}$ and $U_H$ of this component \citep{Costantini07,Arav07}.
The higher ionization phase, which was only detected in the 
x-rays, contained 3 times larger $\vy{N}{H}$ \citep{Costantini07}. Such
separate ionization components are typical for x-ray warm absorbers, where in
some cases the higher ionization x-ray phase has 10 times or more $\vy{N}{H}$
than the lower ionization UV phase. For example, from Table 3 in \citet{Gabel05b},
in the NGC 3783 outflow the combined column density of the warm absorber
is 10-20 times higher than that detected in the UV components. It is clear from Equation \eqref{energetics} that missing
90\%
of the outflowing column density will severely underestimate the $\dot{M}$ and
$\dot{E}_k$ of the outflow.

Second, to determine $R$ we need to measure the $\vy{n}{H}$ of the outflow. 
This necessitates analyzing troughs from excited or metastable 
levels, combined with resonance troughs from the same ion (see \S~5). 
Most quasar outflows show troughs  from only triply or higher ionized species (with the exception of \hi).
Lower ionization species are  observed in only about 10\% of the outflows
\citep{Dai12}.
In this context, it is useful to divide the rest-wavelength UV band into three regions: \\
1) $\lambda>1215$: longward of \Lya, contamination with intervening (mainly IGM) absorption features is small, 
and this is one reason why historically most spectroscopic studies of quasar outflows concentrated on this region.
In this band, only a few ions have transitions from excited states and these are all from singly ionized species (\cii,
\siII\ and \feii). Most of our previous determinations of $R$, $\dot{M}$ and $\dot{E}_k$ (as
well as those of other groups, e.g., \citen{Hamann01})
came from these singly ionized species. Extrapolating from results based on singly
ionized species to the majority of higher ionization outflows introduces two
significant uncertainties for the inferred contribution of all outflows to AGN
feedback: a) pure high ionization outflows may be at a different distance scale
than those that show singly ionized species. b) $\Omega$ for the lower
ionization outflows may be smaller than that of the high ionization outflows simply
as a virtue of the former being detected less frequently. Alternatively, \citet{Hall02,Dunn10b}
argue that $\Omega$ for both forms of outflows are
the same and that the rarity of lower ionization outflows is due to line of sight
effects. It is therefore important to directly obtain $\dot{M}$ and
$\dot{E}_k$  of high ionization
outflows in order to reduce the systematic uncertainty in the average $\Omega$. \\
2) $1000<\lambda<1215$: this spectral region contains excited transitions from \ciii* (1175\AA) and \siv* (1072\AA).
In high-luminosity, high redshift quasars, troughs from these transitions can be severely contaminated with \Lya\ forest absorption features.
In addition, the \ciii* multiplet consists of 6 transitions that span a total of 1.5 \AA\ (less than 400 \kms), 
therefore absorption features from these transitions often self-blend.  Despite these difficulties, the importance of 
these diagnostics to the analysis of quasar outflows, led us to pursue such troughs whenever possible \citep{Dunn12,Borguet12b,Borguet13}.

3) As we show in the next subsection and in Figure 1, the rest wavelength region between $500-1000$ \AA\ contains 
a far richer and more powerful set of outflow diagnostics; even more so when combined with the already available coverage of the 
$\lambda>1000$ \AA\ spectral region.

\subsection{Powerful Outflows Diagnostics in the Far UV (500--1000\AA)}

An in-depth physical analysis of quasar outflows necessitates a careful treatment of all the five steps 
mentioned in \S~1. The starting point is the spectrum on hand with its instrumental quality 
(S/N and spectral resolution) and the diagnostic power of the detected outflow troughs.
For the ubiquitous high ionization outflows (those that show troughs from only triply or higher ionized species), 
we noted in \S~1 the dearth of useful troughs longwards of 1000 \AA.
The situation is quite different for the 500--1000 \AA\ band. In that spectral range we encounter 
many ionic transitions from high-ionization species which are detected as troughs in most quasar outflows. These troughs supply powerful combined diagnostics 
 that have the potential to revolutionize our understanding 
of the connection between the outflow and the host galaxy and its surroundings.
Figure 1 gives a visual representation to the more important ionic transitions available for quasar outflows.

\subsubsection{Sensitivity to the warm absorber phase of the outflow}
Much of what we have learned about AGN
outflows come from the lithium like iso-sequence since they give rise to
observed doublet troughs from \civ\ , \nv\ and \ovi\ at
$\lambda_{rest}\gtorder1000$ \AA . The same iso-sequence has similar doublet lines
from progressively higher-ionization abundant species shortwards of
$\lambda_{rest}=1000$ \AA: \neviii~$\lambda\lambda$~770.409,780.324, \naix~$\lambda\lambda$~681.719,694.146,
\mgx~$\lambda\lambda$609.793,624.941 and
\sixii~$\lambda\lambda$~499.406,520.665.
As we show in Section 6, \neviii\ and \mgx\ are sensitive to the same ionization parameter as 
the bulk of the material reported in detailed x-ray warm-absorber analysis \citep[e.g.][]{Gabel05a,Costantini07}.
For HE0238--1904, measurements of the doublets associated with these two ions yields enough information on the high ionization phase
to show that in this object it carries $\sim$100 times more mass and energy than the
lower ionization phase. 

\subsubsection{Eliminating the major uncertainty regarding $\Omega$}

Shortwards of $\lambda_{rest}=1000$ \AA\ there
are several instances of resonance and excited troughs from the same
high-ionization species, which produce outflow absorption troughs
(e.g., \niii/\niii*$\lambda\lambda989.799,991.577$;
\oiv/\oiv*~$\lambda\lambda787.711,790.199$ and \oiv/\oiv*~$\lambda\lambda609.829,608.397$, more are displayed in Figure~1).
As we show in Section 7, for HE0238--1904, measurements of
\oiv/\oiv*~$\lambda\lambda787.711,790.199$ yield a distance of $R=3000$ pc for
the outflow.  Since high-ionization outflows are the large majority of observed
quasar outflows, we can take their detection frequency \citep[$\simeq$50\%][and especially \citen{Muzahid13}]{Ganguly08,Dai08}
as a robust statistical average for
the $\Omega$ of these outflows. This, removes the second major uncertainty
discussed in the introduction. 

\subsubsection{Separating abundances and photoionization effects}
 
To find the total column density ($\vy{N}{H}$) and the ionization parameter ($U_H$) of the absorbing material,
we use photoionization models.
The unknown chemical abundances of the absorbing gas introduces a significant 
uncertainty in these models when we use, as input, ionic column densities of different elements. 
For example, the same measured column densities for \hi\ and \oiv\ can yield a solution with very different $\vy{N}{H}$ and $U_H$, 
for modest changes in the oxygen abundance, as the changes in $\vy{N}{H}$ and $U_H$ are highly non-linear with the assumed abundances 
(as can be seen in Figure~8). 
This uncertainty can be largely removed by using column densities of more than one ion from the same element.

Unlike the longer wavelength band, the rest wavelength
between $500-1000$ \AA\ contains several instances of resonance troughs from
different ions of the same element. Such occurences greatly reduce the errors in
determining the photoionization equilibrium due to chemical abundance
uncertainties \citep{Arav07} and dust depletion \citep{Dunn10a}. 
For example, in the region $\lambda_{rest}>1000$\AA\ we only detect troughs from one oxygen ion (\ovi~$\lambda\lambda$~1031.926,1037.617)
and usually obtain upper limits for \oi~$\lambda$~1302.17. In contrast, the rest 500--1000\AA\ spectral region covers transitions from 
\oii, \oiii, \oiv\ and \ov.  Therefore, a full spectral coverage for $\lambda_{rest}>500$\AA\ yields information from six oxygen ions.
Similar situations occur for carbon, nitrogen, neon and sulfur. 
As we show
in Section~6, for HE0238--1904 measurements, troughs from \oiv\ and
\ovi\ are crucial for deciphering the ionization structure of the outflow.

\subsubsection{Measuring reliable ionic column densities}

Reliable measurements of the absorption ionic column densities ($N_{ion}$) in the troughs are crucial for determining
almost every physical aspect of the outflows: ionization equilibrium and abundances, number
density, distance, mass flow rate and kinetic luminosity. Our group \citep{Arav97,Arav99a,Arav99b,deKool01,Arav01,Arav01b,Arav02,Arav03,Scott04,Gabel05a}
and others \citep{Barlow97a,Telfer98,Churchill99,Ganguly99}
have shown that outflow troughs often exhibit non-black saturation. Therefore,
traditional absorption method techniques (apparent optical depth, equivalent width and curve of growth) often severely underestimate the 
true $N_{ion}$. A major improvement is achieved by using resonance doublet lines that allow fitting more sophisticated models, 
mainly a pure partial covering model, or an inhomogeneous model \citep[e.g.][]{Arav05}.  However, both of these models have two free parameters, 
and we can usually find a perfect fit for either model given an unblended doublet trough.  
In a few cases we were able to fit  more than two lines from the same ion and thus determine which model gives a better fit to the data.
To date, the few published analyses to this end \citep[e.g.][]{Arav08,Borguet12a} used low ionization species (mainly \feii\ and \siII).

Extending the spectral coverage to $500-1100$ \AA\ yield several ionic species with multiple resonance transitions.
The most important of these is the \hi\ Lyman series. It serves both as a very sensitive measure of the true $N_H$ as well as the best probe 
for the chemical abundances of the outflow (when compared with $N_{ion}$ of heavier elements). Other useful examples are:
\ovi\ 787.7105\AA, 608.3968\AA, 554.0756\AA, 553.3293\AA\ with more than a spread of 3 in oscillator strength ($f$);
\siv\ 1062.6640\AA, 809.6556\AA, 744.9045\AA, 657.3187\AA\ with more than a spread of 20 in $f$.

\begin{figure}
\includegraphics[scale=0.7,angle=0]{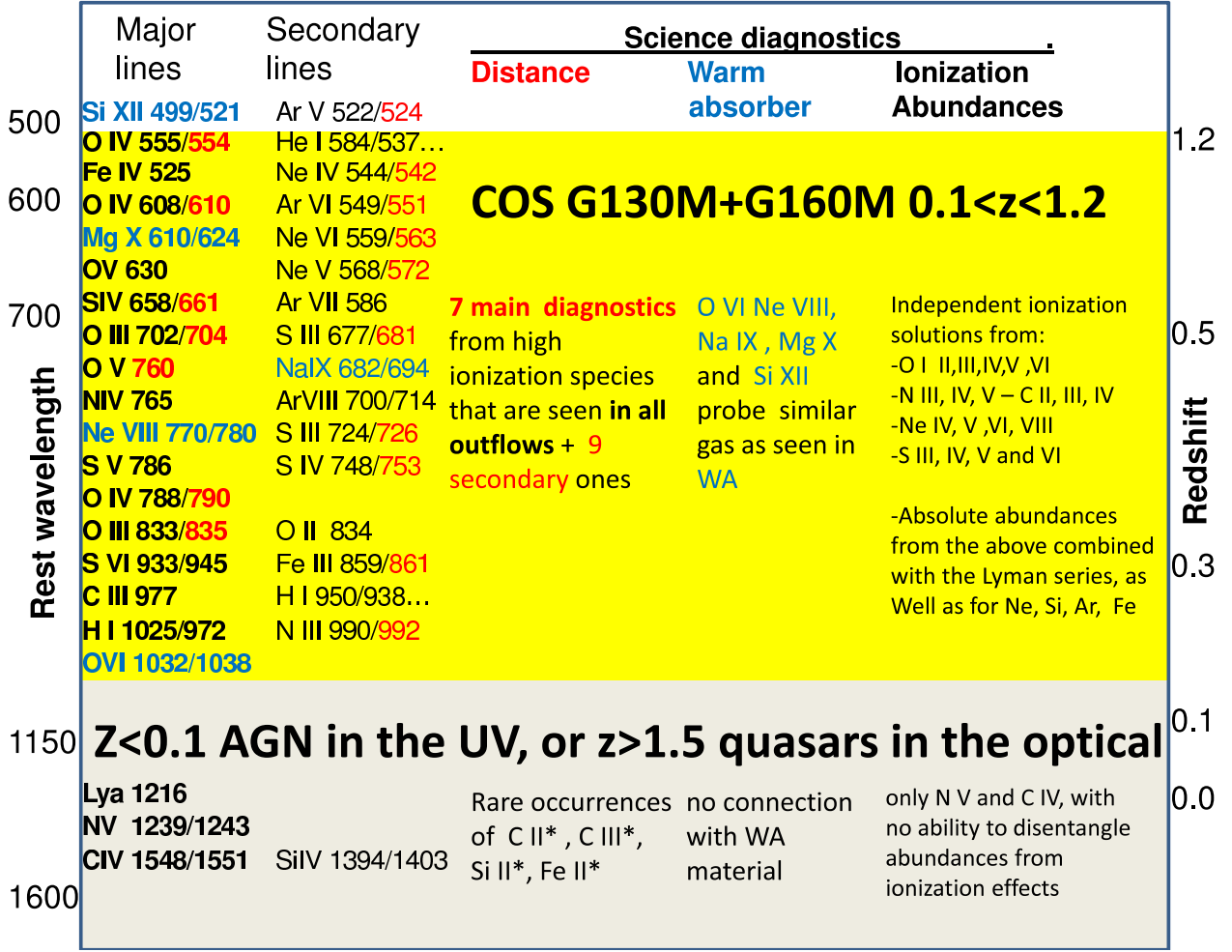}
\caption{With
20 times higher throughput in the FUV than STIS Echelle, COS G130M and G160M
gratings can observe quasars at the 0.1$<$ z$<$1.3 range with sufficient 
S/N and resolution to extract outflow science. This opens up the
spectral-diagnostics-rich 500--1050\AA\ rest-frame region 
 (yellow band).  The left two columns show the lines observed as
absorption troughs in quasar outflows. In the right three columns we show the use of these
lines as physical diagnostics in the
three main analysis areas: a)
distance to the outflows is determined by using absorption from 
excited states (noted in red); b) direct connection to the x-ray warm
absorber  is achieved by observing very high ionization
lines  (noted in blue); c)  ionization equilibrium and
abundances effects are  separated their  by analyzing absorption from multiple 
ions from the same element and measuring several lines from the Lyman series. 
}
\label{diagnostics}
\end{figure}

\pagebreak
\section{DATA OVERVIEW}

\subsection{Observations and data reduction}

The quasar HE\,0238$-$1904 was observed with HST/COS \citep[see][for on-orbit performance]{Osterman10} 
on 2009 December 31.
Spectra of the target were obtained using the Primary Science Aperture (PSA) in both
the G130M ($1135<\lambda<1480$ \AA) and G160M ($1400<\lambda<1795$
\AA) medium-resolution ($R=\lambda/\Delta \lambda\approx 18,000$)
gratings as part of the COS Guaranteed Time Observations (GTO) program
(PI: Green, PID: 11541) and totaled 6451 s and 7487 s in the G130M and G160M
gratings, respectively. Both sets of observations used multiple
grating central settings in order to provide a continuous spectral coverage
across the entire COS far-UV band as well as dither instrumental features
in wavelength space.

Each dataset is reduced with the COS calibration pipeline {\sc CALCOS} v2.11f.
Flat-fielding, alignment and co-addition of the processed exposures is then
carried out using IDL routines developed by the COS GTO team specifically
for COS FUV data\footnote{IDL routines available at {\tt
http://casa.colorado.edu/$\sim$danforth/costools.html}} and described
in \citet{Danforth10}. Briefly, each exposure is corrected for
narrow, $\sim15\%$-opaque, shadows from repellor grid wires. The local
exposure time in these regions was reduced to give them less weight in
the final co-addition. Similarly, exposure times for data at the edges
of the detectors was de-weighted.  With multiple central wavelength
settings per grating, any residual instrumental artifacts from
grid-wire shadows and detector boundaries have negligible effect on
the final spectrum.

\begin{figure}
\includegraphics[scale=0.7,angle=90]{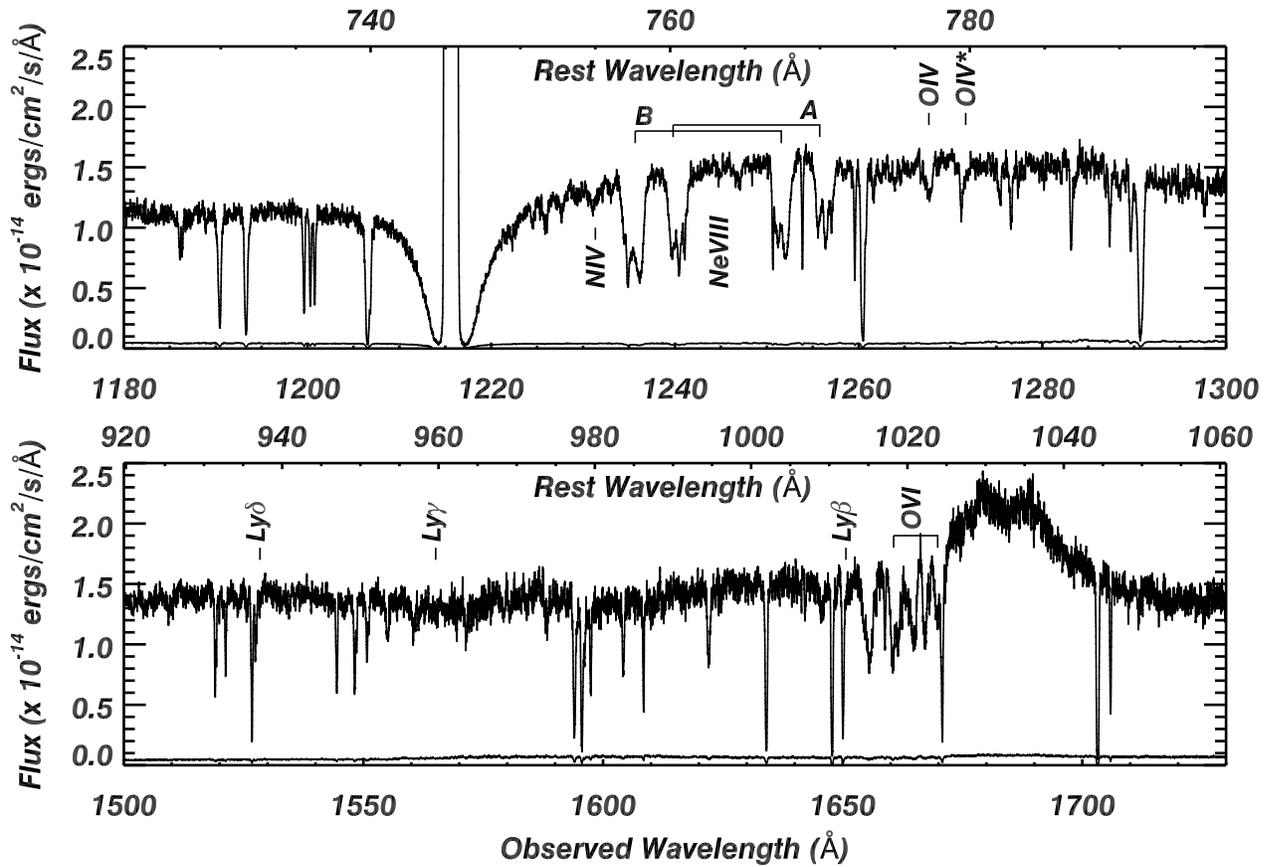}
\caption{This figure presents over $80$\% of the fully reduced HE0238-1904 spectrum observed
in December 2009 with HST/COS. The position of the main intrinsic absorption features are labeled.}
\label{fullspec}
\end{figure}

Next, strong ISM features in each exposure are aligned via
cross-correlation and interpolated onto a common wavelength scale.
The wavelength shifts are typically on the order of a resolution
element ($\sim15$~\kms) or less. The co-added flux at
each wavelength is taken to be the exposure-weighted mean of flux in
each exposure. To quantify the quality of the combined data, we
identify line-free continuum regions at various wavelengths, smooth
the data by the seven-pixel resolution element and define $S/N\equiv
F/\Delta F \approx 20-25$ in the data,where $\Delta F$ is
the standard deviation of the flux. In Fig.~\ref{fullspec} we
present sections of the fully reduced COS spectrum along with the identification
of intrinsic absorption troughs.

%\begin{figure}
%  \includegraphics[scale=0.7,angle=90]{he0238_fullspec}
% \caption{This figure presents over $80$\% of the fully reduced HE0238-1904 spectrum observed
%in December 2009 with HST/COS. The position of the main intrinsic absorption features are labeled.}
% \label{fullspec} 
%\end{figure}

\subsection{Unabsorbed emission model}\label{sec:emissionModel}

 Determining the column density associated with each ion requires the knowledge
of the unabsorbed emission model $F_0(v)$. The typical AGN emission is comprised of
three main sources: a continuum emission, a broad emission line component (BEL) and
a narrow emission line component (NEL).  We model the de-reddened ($R_V=3.1$,
$E(B-V)=0.032$ \citep{Schlegel98}) continuum emission of HE0238-1904 using
a single power law of the form $F_{(\lambda)} = F_{1100} (\lambda/1100)^{\alpha}$.
A $\chi^2$ minimization of the model over regions free of known emission/ absorption lines gives
an overall good fit with parameters $F_{1100}=1.714 \times 10^{-14} \pm 0.008 \times 10^{-14}$ and
$\alpha=-0.234 \pm 0.043$.

The emission line profiles in the HE0238-1904 spectrum are generally smooth and shallow.
We fit the prominent \ovi\ emission using two broad gaussians of $FWHM \sim 4000$ and $\sim 12000~ \kms$
for the BEL and a single NEL with $FWHM \sim 700~ \kms$ centered on each line of the doublet.
The remaining weaker emission features are modelled by a spline fit. In Fig.~\ref{em_compo_he0238}, we present
the unabsorbed emission model we constructed over the \ovi\ and \neviii\ regions of HE0238-1904.

\section{ABSORPTION TROUGHS CHARACTERISATION}
%\section{Column density determination}
\label{coldens_absmo}

\subsection{Identification of intrinsic absorption features}
\label{intrinsic_abs}

The COS FUV spectrum of HE0238-1904 ($z=0.6309$) displays a variety of absorption features
in part due to the interstellar medium in our galaxy (ISM); to a collection of intervening Lyman-alpha
systems (IGM); and absorption features that are directly related to an
intrinsic outflow from the quasar.

\begin{figure}
  \includegraphics[scale=0.5,angle=90]{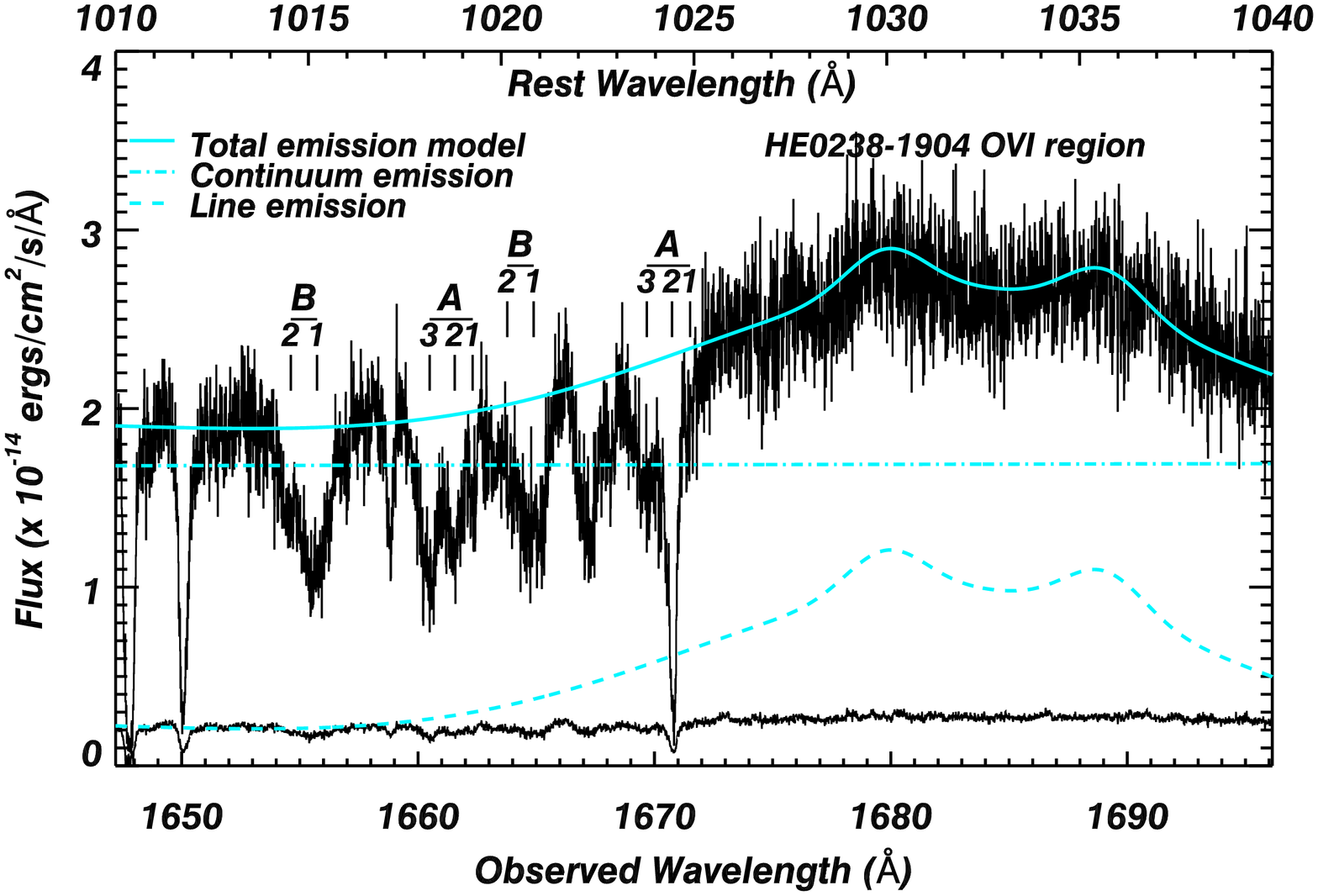}\\
  \includegraphics[scale=0.5,angle=90]{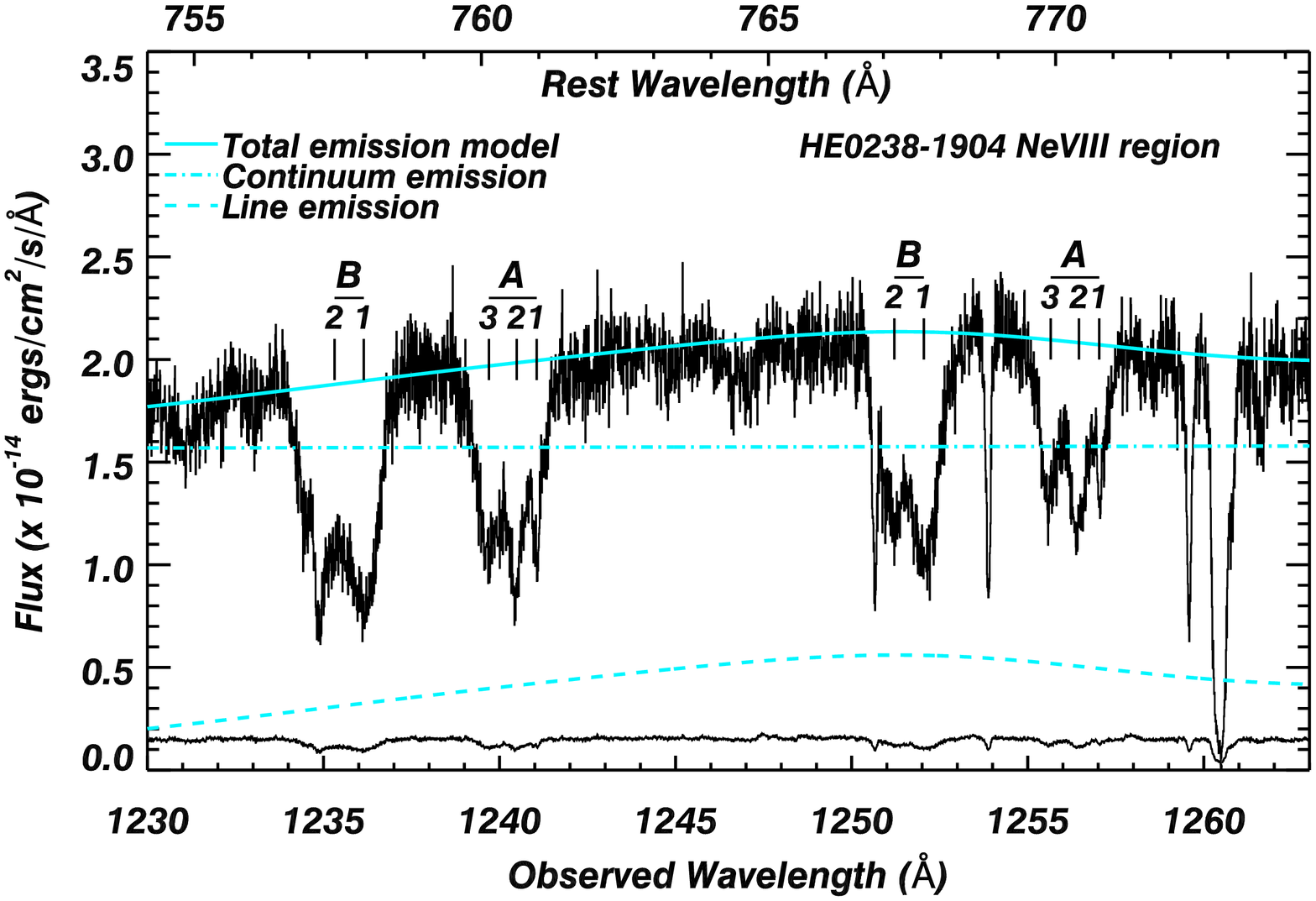}\\
 \caption{Detail of the unabsorbed emission model for the \ovi\ and \neviii\ emission line region on the dereddened spectrum of HE0238-1904 (see text). The total emission model is
  plotted as a solid line on top of the data, the continuum contribution is plotted in dotted-dashed line and the line emission is plotted in
  dashed line. We also indicate the position of the absorbing systems identified from comparing the \neviii\ and \ovi\ line profiles.
  The error spectrum is also shown at the bottom.}
 \label{em_compo_he0238} 
\end{figure}

Using the line profile of the high ionization \neviii\ and \ovi\ doublets as templates, we identify two main absorbing systems
($A$ and $B$), blueshifted at velocities $v_A \sim -3850 ~ \kms$ and
$v_B \sim -5000 ~ \kms$, both exhibiting a width $\Delta v \sim 500 ~\kms$.
The resolution of COS allows us to resolve each absorption system into several narrower components.
System $A$ exhibits three main subcomponents ($A1, A2, A3$) located at velocities $v_{A1} \sim -3698 ~ \kms$,
$v_{A2} \sim -3835 ~ \kms$, $v_{A3} \sim -4026 ~ \kms$ while system $B$ essentially displays
two subcomponents ($B1, B2$) at velocities $v_{B1} \sim -4887 ~ \kms$, $v_{B2} \sim -5088 ~ \kms$ (see Fig.~\ref{em_compo_he0238}).

Using that kinematic structure template we identify absorption
features associated with the outflow in lower ionization species such as \niv, \oiv, and \svi\ as well as lines from the Lyman series (Ly$\beta$, Ly$\gamma$
and Ly$\delta$). The absorption detected in these ions are matching the kinematic
structure of subcomponents $A3$ and $B1$, while displaying shallower troughs than the high ionization \neviii\ and \ovi\ templates.
An absorption feature associated with the $\lambda 790.199$ transition of excited
 \oiv\ (\oiv*) is detected in outflow component
$A3$, while its blue wing is affected by a blend with an intervening Ly$\epsilon$ line.

The line of sight towards HE0238-1904 shows numerous intervening Lyman systems
occasioning mild to severe blending of the diagnostic lines we use to characterize the
properties of the intrinsic outflow. Using the strong Ly$\alpha$
and Ly$\beta$ transitions, we identify at least 30 intervening Lyman systems
in that direction (the analysis of these systems will be given in a future paper:
Danforth et al. 2013). We detail the blends affecting each
individual intrinsic absorption line and how we treat them in the next sections. The case
of the \oiv* line is of particular interest in our study, given that it allows us
to estimate the density and hence the distance to the outflow.

\begin{figure}
  \includegraphics[scale=0.7,angle=90]{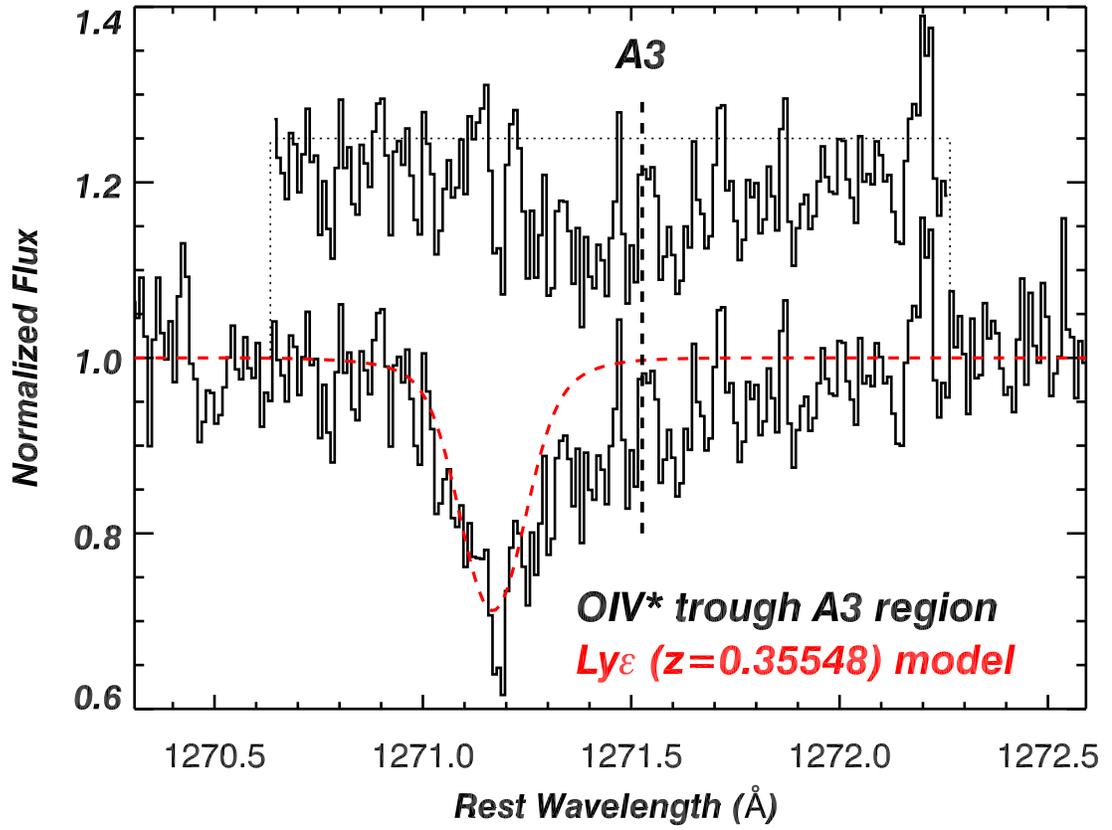}\\
 \caption{Deblending of component $A3$ of the intrinsic \oiv* $\lambda 790.199$ by dividing out
  the Ly$\epsilon$ absorption model from the strong intervening system at $z=0.35548$ (see text). The
  corrected normalized \oiv* profile, plotted here for clarity with a shifted continuum, matches the kinematic structure of other ions associated with the outflow.}
 \label{oiv_lye_corr} 
\end{figure}

Component $A3$ of the \oiv* absorption line associated with the outflow of HE0238-1904
is observed at 1271.2 \AA, and is blended with Ly$\epsilon$ absorption from a strong
IGM system at $z=0.35548$ lying on its blue wing. Thanks to the broad spectral coverage and good data
quality of the COS observations, we can model the \hi\ absorption out fairly accurately.
We measured the equivalent widths of \hi\ absorption in the $z=0.35548$ system using the
Ly$\alpha$-Ly$\zeta$ transitions (excepting Ly$\epsilon$). These equivalent widths were well-fit
by a single curve-of-growth with parameters $\log\,N_{\rm HI}=14.95\pm0.05$ and $b_{\rm HI}=21\pm1$~km~s$^{-1}$. 
The constructed Ly$\epsilon$ trough model $M_{\rm{Ly} \epsilon}(\lambda)=e^{-\tau_{\rm{Ly} \epsilon}(\lambda)}$ was then divided out following the procedure
$I_{\rm{IGM-corrected}}(\lambda) = I(\lambda)/M_{\rm{Ly} \epsilon}(\lambda)$, revealing the \oiv* trough whose
kinematic structure in trough $A3$ matches the one observed in the other ions of the outflow
(see Fig.~\ref{oiv_lye_corr})

\subsection{Inhomogeneities of the absorber}

 The column density associated with a given ion
 as a function of the radial velocity $v$ is defined as:
    \begin{equation}
    \label{eqcoldens}
     {N_{ion}(v) = \frac{3.8 \times 10^{14}}{f_j \lambda_j}  <\tau_j(v)> ~~(\textrm{cm}^{-2}\textrm{ km}^{-1}\textrm{ s})}
   \end{equation} 
  where $f_j$, $ \lambda_j$ and $<\tau_j(v)>$ are respectively the oscillator strength, the rest wavelength
 and the average optical depth across the emission source of the line $j$ for which the optical depth solution
 is derived \citep[see][]{Edmonds11}. The optical depth solution across a trough is found for a given ion by assuming an
 absorber model. As shown in \citet{Edmonds11}, the major uncertainty on the derived column densities
 comes from the choice of absorption model. In this study we investigate the outflow properties using column densities
 derived from three common absorber models.

 Assuming a single, homogeneous emission source of intensity $F_0$, the simplest absorber model
is the one where a homogeneous absorber parameterized by a single optical depth fully covers the photon source.
 In that case, known as the apparent optical depth scenario (AOD), the optical depth of a line $j$
 as a function of the radial velocity $v$ in the trough is simply derived by the inversion of
 the Beer-Lambert law : $\tau_j(v)=-ln(F_j(v)/F_0(v))$, where $F_j(v)$ is the observed intensity
 of the line.

 Early studies of AGN outflows pointed out the inadequacy of such an absorber
 model, specifically its inability to account for the observed departure of measured optical depth ratio between
 the components of typical doublet lines from the expected laboratory
 line strength ratio $R = \lambda_i f_i /\lambda_j f_j$. Two parameter absorber
 models have been developed to explain such discrepancies.
 The partial covering model \citep[e.g.][]{Hamann97b,Arav99b,Arav02,Arav05} assumes
 that only a fraction $C$ of the emission source is covered by absorbing material with constant optical depth $\tau$.
 In that case, the intensity observed for a line $j$ of a chosen ion can be expressed as
 \begin{equation}
 \label{eqcov}
 { F_j(v) = F_0(v) (1+C(v)*(e^{-\tau{j}(v)}-1)).}
\end{equation}

 Our third choice are inhomogeneous absorber models.
 In that scenario, the emission source is totally covered by a smooth distribution of
 absorbing material across its spatial dimension $x$. Assuming the typical power law
 distribution of the optical depth $\tau(x) = \tau_{max} x^{a}$ \citep{deKool02,Arav05,Arav08},
 the observed intensity observed for a line $j$ of a chosen ion is given by
  \begin{equation}
    \label{eqpow}
    {F_j(v) = F_0(v)  \int^1_0  e^{-\tau_{max,j}(v) x^{ a(v)} } dx }
  \end{equation}

 Once the line profiles have been binned on a common velocity scale (we choose a resolution $dv =20~\kms$,
 slightly lower than the resolution of COS), a velocity dependent solution can be obtained for the couple of parameters
 $(C,\tau_{j})$ or $(a,\tau_{max})$ of both absorber models as long as one observes
 at least two lines from a given ion, sharing the same lower energy level.
 Once the velocity dependent solution is computed, the corresponding column
 density is derived using Equation \ref{eqcoldens} where $<\tau_j(v)> = C_{ion}(v) \tau_j(v)$ for
 the partial covering model and $<\tau_j(v)>= \tau_{max,j}(v)/(a_{ion}(v) + 1)$ for the power law distribution.
 Note that the AOD solution can be computed for any line (singlet or multiplet), without further
 assumption on the model, but will essentially give a lower limit on the column density when
the expected line strength ratio observed is different from the laboratory value.

\section{COLUMN DENSITY MEASUREMENTS}

  As mentioned in Section \ref{intrinsic_abs}, the line of sight towards
 the quasar HE0238-1904, in addition to the usual interstellar medium lines,
 has many intervening Lyman systems affecting our
 ability to derive accurate ionic column density of lines associated
 with the intrinsic outflow, if not properly taken into account.
 In the following subsections we detail individually the blends affecting each intrinsic
 absorption troughs and present the analysis we performed in order to estimate
 the column density of each ion.

 The compilation of measurements is presented in Table 1 in which the ionic column densities
 have been computed on four different sections of the main troughs $A$ and $B$. The first three
 trough sections are related to absorption system $A$ in which a clear signature of an excited state of
 \oiv\ is detected in at least subtrough $A3$. Though the detection of \oiv\ and \oiv*  in $A3$ allows us to derive a distance to
 that component, the bulk of the column density comes from the total trough $A$. While the high velocity
 $v_A \sim -3850 ~ \kms$ and narrowness $\Delta v \sim 500 ~ \kms$ of the system suggests a 
 physical connection between its subcomponents $A1$, $A2$ and $A3$, we further test the validity
 of that assumption by comparing the ionization solution for different sections of the main trough.
 The first integration range considered: $v_{A3} \in [-4050,-3930]~\kms$ is mainly centered on trough $A3$,
 but where the lower integration limit is chosen to avoid the blue wing of \oiv*, giving a
 measurement virtually independent of the correction of the Ly$\epsilon$ line blend. The second range 
 consists of the remaining trough: $v_{A1+A2} \in [-3930,-3610]~\kms$, essentially comprising
 components $A1$+$A2$. The third range integrates the column density over the whole system $A$:
 $v_A \in [-4170,-3610]~\kms$. The last region, $v_B \in [-5270, -4650]~\kms$, simply
 integrate the column density over the whole higher velocity system $B$.

In Table 1 we compare the estimated column densities derived in the presented
analysis and the values reported in the analysis of the same absorber in \citet{Muzahid12}. We summed
the column densities reported in their paper and multiplied its value by the average covering of each species
from their Table~1 in order to be consistent with the average values we derive (see Section~\ref{coldens_absmo}).

\begin{figure}
  \includegraphics[scale=1.2,angle=90]{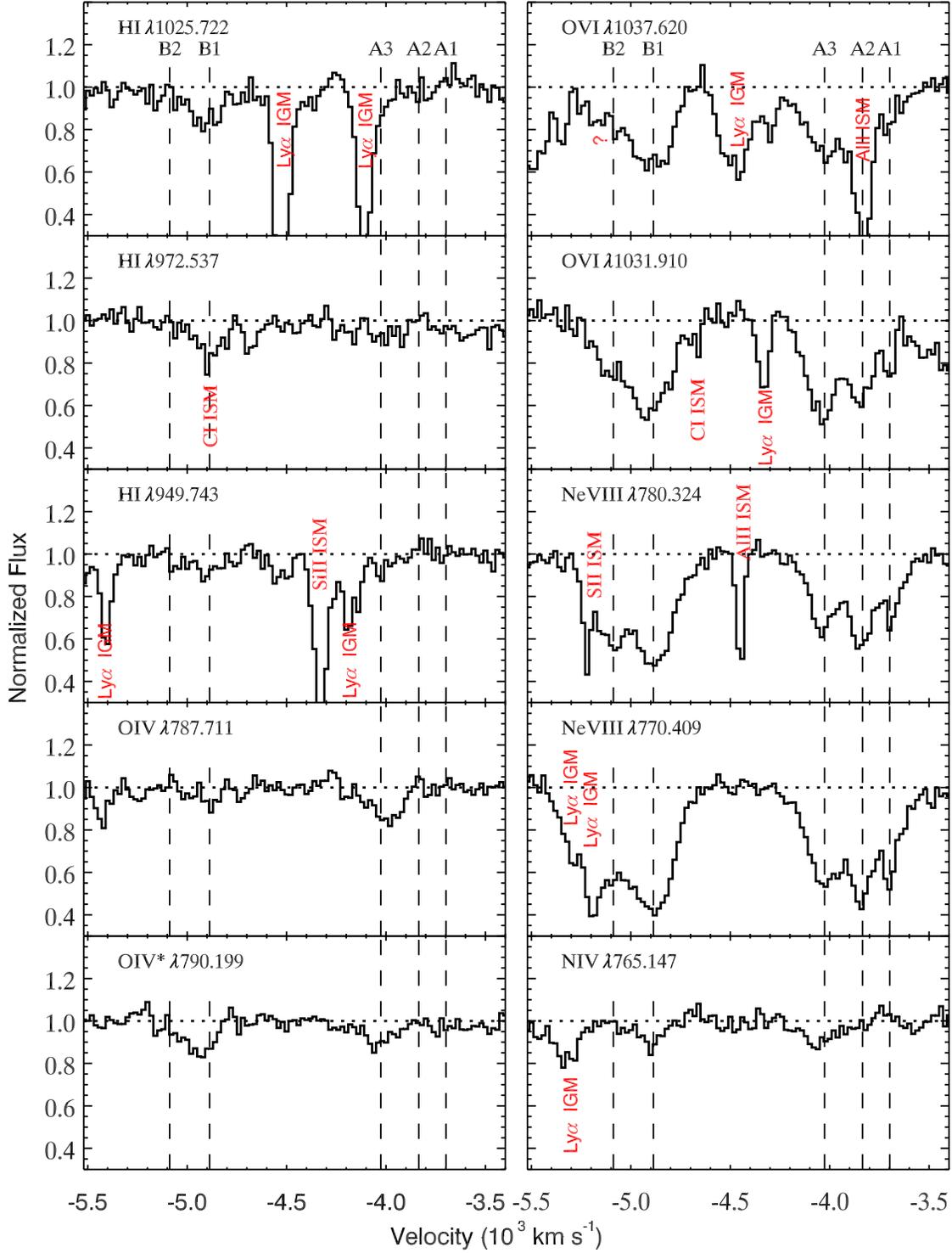}\\
 \caption{This figure presents the line profiles associated with ions identified in the outflow.
 the dashed line indicate the position of the outflow components. We label the blends affecting
 each line. The ``?'' in the top right panel corresponds to a probable unknown blend (see Section~\ref{zeosix})}
 \label{all_line} 
\end{figure}

\subsection{The \neviii\ troughs}

We do not identify any blending in both components of the \neviii\ doublet in system $A$
which allows us to derive a good column density solution over each subregion ($v_A$, $v_{A1+A2}$ and $v_{A3}$) of the trough
using the three absorber models described in Section~\ref{coldens_absmo}.
In system $B$, the blue wing of \neviii\ $\lambda ~780.324$ line is blended between $1250-1251$ \AA~ by
the ISM \sii\ $\lambda ~1250.578$ line and also by a weak Ly$\epsilon$ IGM ($z=0.3343$) line near $1251.2$ \AA.
The blue wing of the blue component of \neviii\ $\lambda ~770.409$ is affected by a blend with two Ly$\alpha$ IGM lines ($z=0.0155$ and $z=0.0158$)
as well as a weak blend by a Ly$\gamma$ IGM ($z=0.2704$) line near $1235$ \AA. Given the
different blending, a crude estimate of the \neviii\ column density is computed by using the three
absorber solutions in the blend free region $v \in [-4750,-3650]~\kms$ and adding to it a lower limit on
the column density that can be hidden in the blended region by scaling the
optical depth solution of the non-blended \ovi\ $\lambda 1031.910$ line template
to the position of the \neviii\ $\lambda ~780.324$ line.

\subsection{The \ovi\ troughs}
\label{zeosix}

The central part and the red wing of the \ovi\ $\lambda 1037.620$ line in trough $A$
are blended between $1670-1672$ \AA~ by the strong ISM \alii\ $\lambda 1670.787$ line.
No blend is identified on the \ovi\ $\lambda 1031.910$ trough. We checked the portion
of the \ovi\ $\lambda 1037.620$ affected by the blend by overplotting the unblended and
non-saturated ISM \siII\ $\lambda 1193.290$ (with an ionization potential close to \alii) at the rest position of the \alii\ $\lambda 1670.790$
line, revealing that while the $v_{A1+A2}$ region of \ovi\ $\lambda 1037.620$ is the most affected section of
the trough, the blend may still significantly affect the red part of the $v_{A3}$ region.
Taking that observation into account, we derive a lower limit on the ionic column density
in region $v_{A3}$ using the AOD method on \ovi\ $\lambda 1031.910$. We estimate an upper
limit for the same region by scaling the column density derived using the partial covering method
over the unblended section of the region $v_{A3}$ (ie. $v \in [-4050, -3970]~\kms$) to the
entire $v_{A3}$ trough width. In region $v_{A1+A2}$ the strong blend of \ovi\ $\lambda 1037.620$
only allow us to derive a lower limit on the column density by using the AOD method on
the unblended \ovi\ $\lambda 1031.910$ line.

Except for a weak blend by \ci\ $\lambda 1656.929$ close to the red wing (but outside the $v_B$ range) of \ovi\ $\lambda 1031.910$,
we do not a-priori identify any major blend in system $B$. However, the comparison of the residual
flux in \ovi\ $\lambda 1031.910$ and \ovi\ $\lambda 1037.620$ strongly suggests a weak
blend around 5200 km s$^{-1}$, since the residual flux in that component
is in overall lower that the residual flux in the stronger \ovi\ $\lambda 1031.910$ line.
Taking that observation into account, we obtain an estimate of the column density
using the three absorber model over the unblended section of the trough ($v \in [-4730, -5170]~\kms$)
and adding the AOD contribution determined from the \ovi\ $\lambda 1031.910$ line in the section where the unidentified
blend affects \ovi\ $\lambda 1037.620$ ($v \in [-5170, -5270]~\kms$). Note that
in this first section of trough $B$, the power law method gives a larger amount of
column density which is an artifact due to the inability of that absorber model to
account for shallow, saturated ($I_{\lambda 1031.910} \sim I_{\lambda 1037.620}$) troughs \citep[see][]{Arav05}.

\subsection{The \hi\ troughs}

We detect absorption troughs associated with neutral hydrogen in
system $A$ in Ly$\beta$, the overall S/N does not allow to
assess the detection of absorption troughs in higher order lines.
In region $v_{A3}$, the blue wing of the weak Ly$\beta$ line is
affected by a blend with a moderate IGM Ly$\alpha$ line at $z=0.35733$.
The absence of unblended higher order line for that IGM system
prevent us to perform a curve-of-growth analysis of the system,
so we modeled the IGM Ly$\alpha$ line by a gaussian profile centered
around the rest Ly$\alpha$ in the $z=0.35733$ frame and divided the model
out. A lower limit of the \hi\ column density in that region $v_{A3}$ is then
derived by scaling the AOD optical depth profile computed from the unblended
\neviii\ $\lambda ~770.409$ template to match the noisy Ly$\beta$ residuals. A conservative upper limit on \hi\
is obtained in that same region by scaling the same template to the noise
level for the Ly$\gamma$ line. In region $v_{A1+A2}$, we estimate the \hi\
column density by using the AOD method on the weak Ly$\beta$ profile.

In system $B$, we identify \hi\ troughs in the Ly$\beta$, Ly$\gamma$ and
possibly Ly$\delta$ lines. The only line suffering from known blending
is Ly$\gamma$ which is blended in the $1560-1561$ \AA ~ region by the \ci\
$\lambda 1560.309$ line. Given the shallowness of both Ly$\gamma$ and \ci\
lines as well as the limited S/N, we do not attempt do deblend these troughs
and only use the Ly$\beta$ and Ly$\delta$ lines to constrain the \hi\ column
density in that region. A conservative lower limit on \hi\ is obtained 
by using the AOD method on the strong Ly$\beta$ line while an upper limit is derived
computing the AOD solution over the shallow Ly$\delta$ line region.

\subsection{The \oiv\ and \oiv* troughs}

A first look at the HE0238-1904 spectrum reveals the presence of absorption troughs associated with \oiv\ $\lambda 787.711$
mainly in components $A3$ and $B1$, while, once corrected for the IGM Ly$\epsilon$ $z=0.35548$ blend (see Sect.~\ref{intrinsic_abs}),
an absorption trough associated with \oiv* $\lambda 790.199$ is detected mainly in component $A3$. Assuming no further blend,
we can estimate the column density associated with these two state by using the AOD method on each of them, leading to
an \oiv\ column of $83.9^{+6.1}_{-5.9}\times 10^{12} \mathrm{cm}^{-2}$ and $34.4^{+5.7}_{-5.1}\times 10^{12} \mathrm{cm}^{-2}$ in region $v_{A3}$ and $v_{B1}$ respectively
and an \oiv* column of $49.5^{+7.1}_{-6.8}\times 10^{12} \mathrm{cm}^{-2}$ in region $v_{A3}$.

\begin{figure}
  \includegraphics[scale=0.35,angle=90]{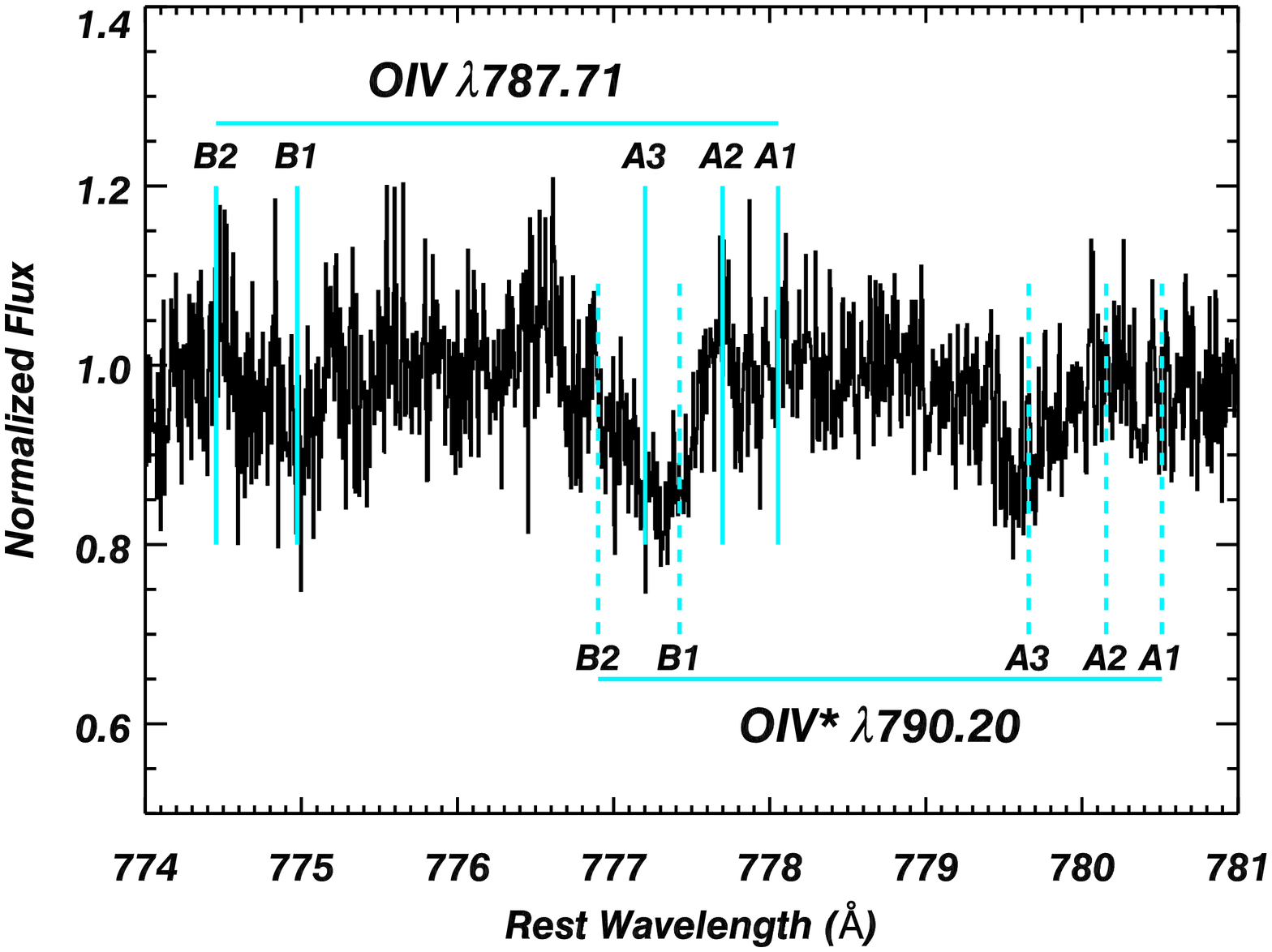}
  \includegraphics[scale=0.35,angle=90]{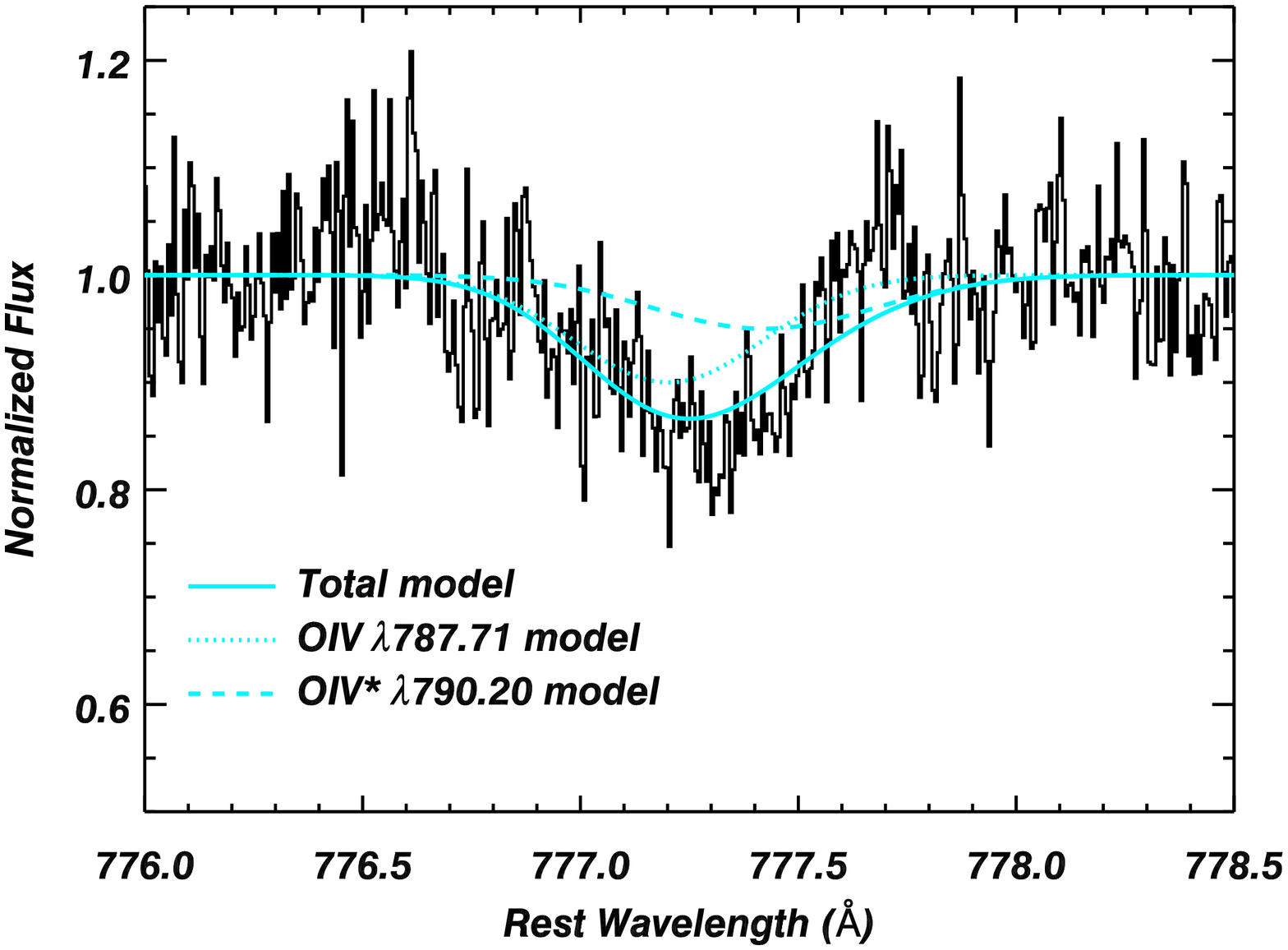}\\
 \caption{In the left panel we show
  the spectral region around the \oiv* and \oiv\ line on which we overplotted the position
  of each subcomponent, suggesting that the trough around the $777.25$ \AA ~ rest-wavelength region is
indeed a blend between \oiv* $B1$ and \oiv\ $A3$ troughs. In the right panel, we show our best fit
to the blend assuming that each subtrough can be modeled by a single gaussian with parameters
derived from the fit of strong unblended troughs of \neviii\ and \ovi\ (see text for details).}
 \label{deblend_oiv} 
\end{figure}

A meticulous look at the \oiv\ $\lambda 787.711$ line profile in trough $A3$ however reveals a kinematic structure slightly
different than the one observed in all the other species, showing an extended asymmetric red wing. That structure can be explained
if that red wing is affected by a blend from an \oiv* $B1$ trough, as suggested in Fig.~\ref{deblend_oiv} in which we plot the expected
position of the \oiv\ and \oiv* components based on the kinematic structure detailed in Section~\ref{intrinsic_abs}.
In order to deblend the \oiv\ $A3$ and \oiv* $B1$ trough, we first modelled the unblended deep but non-saturated components $A3$ and $B1$
of the \neviii\ $\lambda ~770.409$ line by a simple gaussian, as suggested by the general profile of each component.
Using a Levenberg-Marquardt minimization technique, we obtain a good fit to component $A3$ with parameters $v_{cen} = -4026.22 \pm 4.56 ~ \kms$
and $FWHM = 275.27  \pm 17.17 ~ \kms$, where $v_{cen}$ and $FWHM$ are the central position and Full Width at Half Maximum of the best gaussian model.
For component $B1$ we similarly derive $v_{cen} = -4887.04 \pm 3.45 ~ \kms$ and $FWHM = 325.61  \pm 13.46~ \kms$.
Using the derived position and width of these two components, we construct a two component gaussian model of the
\oiv\ and \oiv* blend once again using the same minimization technique. The best fitting model is presented in
Fig.~\ref{deblend_oiv}. Having computed a model of each trough inside the blend, we can correct the blended region in order
to derive the actual column density in \oiv* $B1$ or \oiv\ $A3$ by dividing out the initial normalized line
profile by the \oiv\ or \oiv* model. Applying this technique to the actual spectrum we obtain an AOD measurement
in regions $v_{A3}$, $v_{B}$ as well as $v_{A1+A2}$ for both \oiv\ and \oiv* line, reported in Table~1.

\subsection{Other intrinsic troughs}

As mentioned in Section~\ref{intrinsic_abs}, we detect absorption features associated with
other ionic species in the HE0238-1904 spectrum. The most prominent of them is \niv\ $\lambda 765.147$. No
blends are identified in either system $A$ or $B$ for that line allowing us to derive a reliable AOD solution
for the column density in these two components. A shallow signature is also probably detected in the
strongest line of the \svi\ ion (i.e. \svi\ $\lambda 933.376$). While consistent with the overall noise
level in that spectral range, we compute an upper limit on the \svi\ column density by deriving the AOD
solution over systems $A$ and $B$. 

We also searched for other lower ionization lines like \ciii, \niii\ and \oiii\ in order to better constraint
the photoionization model of each absorber. The only feature identified for these ions is a possible signature
of \oiii\ $\lambda 832.927$ in subcomponent $A3$ which we also identified as a weak Ly$\beta$ associated with the
Lyman intervening system at $z=0.3065$. We derive a conservative upper limit on the column density of each
of these ions by scaling the optical depth template of the unblended \neviii\ $\lambda ~770.409$ in regions $v_{A1+A2}$ $v_{A3}$ and $v_{A}$ and the template of
\ovi\ $\lambda 1031.910$ in region $v_{B}$.

\subsection{FUSE archival data}

HE0238-1904 has been observed by the Far Ultraviolet Spectroscopic Explorer telescope (FUSE) in December 2000, July 2003 and 
September 2004. We downloaded the spectra from the Multimission Archive at Space Telescope (MAST) and processed them with
CalFUSE v3.2.3 \citep{Dixon07}.Given the apparent lack of variability in the \neviii\ absorption troughs located in the
higher S/N part of the spectra, we combined them together in order to produce a coadded spectrum with S/N $\sim [2 - 6]$ per
resolution element (R $\sim 20000$) that covers the $[920-1180]$ \AA\ range of FUSE. Comparison of the \neviii\ troughs with
the one observed with COS show no apparent changes within the lower FUSE S/N.

\begin{figure}
  \includegraphics[scale=1.0,angle=90]{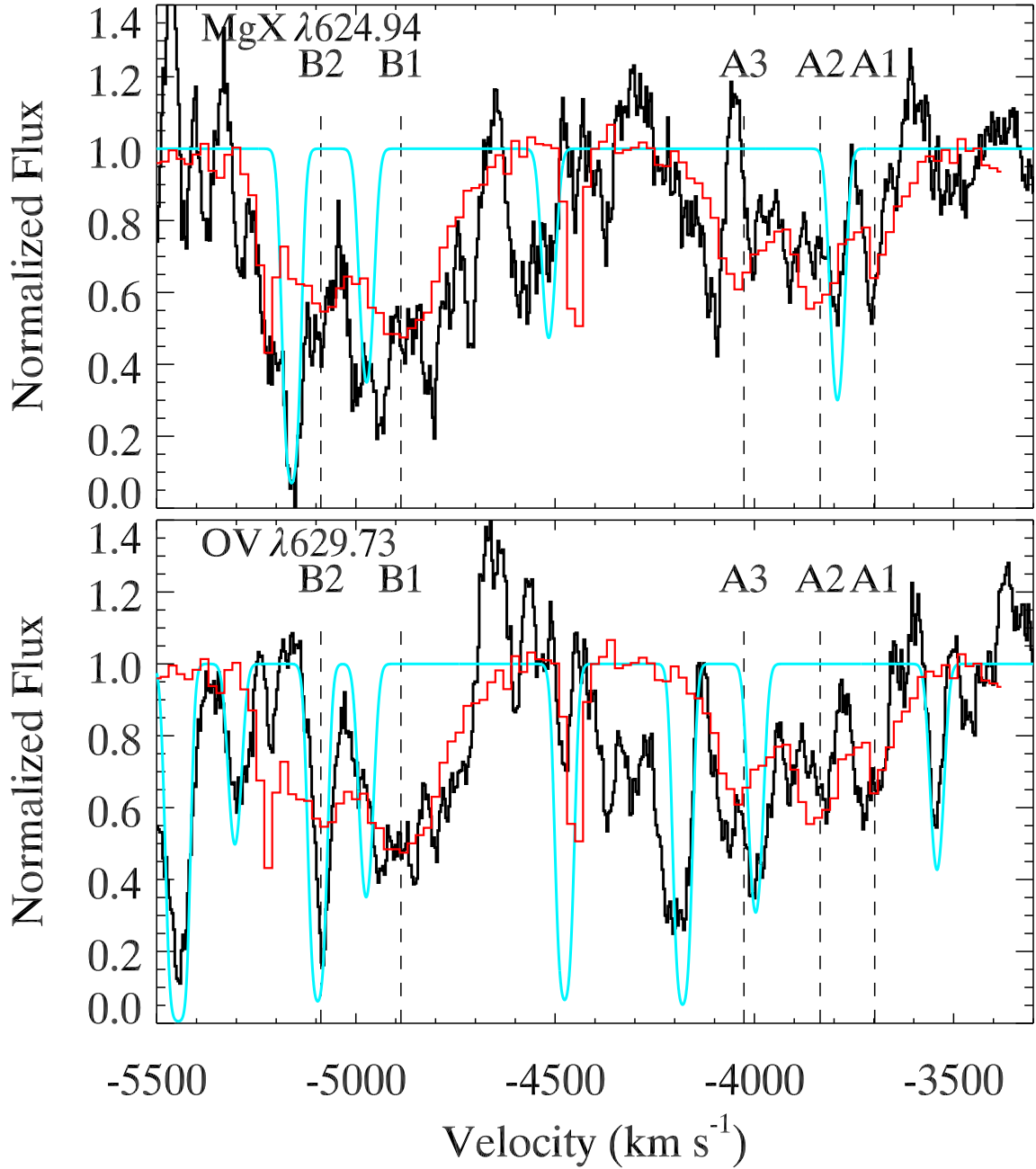}\\
 \caption{FUSE spectrum of HE0238-1904 in the \ov\ and \mgx\ absorption line regions. The FUSE spectrum (in black)
has been smoothed by an 8 pixel boxcar for clarity due to the limited S/N of the observations. In light blue we
display an tentative Galactic $H_{2}$ absorption model considering only the first 6 J levels. In red we overplotted
the \neviii\ $\lambda 780.324$ line profile, helping in the identification of the intrinsic \ov\ and \mgx\ absorption
features.}
 \label{fusetroughs} 
\end{figure}

In addition to the troughs reported in the recent
COS observations, we detect absorption associated with the high ionization \mgx\ $\lambda 624.94$  and \ov $\lambda 629.73$
within the FUV extended FUSE range (see Figure~\ref{fusetroughs}). The limited S/N in the bluer range of FUSE combined with the contamination from Galactic $H_{2}$
absorption lines does not allow us to confirm the detection of the strongest \mgx\ $\lambda 609.79$ absorption troughs.
We estimate the column density present in \mgx\ and \ov\ by scaling the high S/N non-blended COS template of \neviii\ $\lambda 770.41$
and \ovi\ $\lambda 1031.91$ to the observed residual intensities of both species in the FUSE spectra for components in region $TA$ and $TB$ respectively.  
We report the estimated column densities in Table~1 and consider them as lower limits in the remainder of
the analysis.

\begin{deluxetable}{lrrrrr}
\tablecaption{\sc Column densities}
\tablewidth{0pt}
\tablehead{\colhead{Ions}&\colhead{$v_{A3}$}&\colhead{$v_{A}$}&\colhead{$v_{A}$ M12$^{\mathrm{a}}$}&\colhead{$v_{B}$}&\colhead{$v_{B}$ M12$^{\mathrm{a}}$}\\
\colhead{}&\colhead{$(10^{12}~\mathrm{cm}^{-2})$}&\colhead{$(10^{12}~\mathrm{cm}^{-2})$}&\colhead{$(10^{12}~\mathrm{cm}^{-2})$}&\colhead{$(10^{12}~\mathrm{cm}^{-2})$}&\colhead{$(10^{12}~\mathrm{cm}^{-2})$}}
\startdata
 
\hi\ &        $\in [50.9, 183]$&    $\in [92.5, 225]$&        $\leq 315$&    $\in [165, 506]$&    $\in [188,612]$\\
\niv\ &        $6.82^{+1.20}_{-1.13}$&    $19.9^{+2.1}_{-2.1}$&        21&        $12.2^{+1.5}_{-1.4}$&    20\\
\oiv\ &        $64.2^{+6.1}_{-5.9}$&    $98.7^{+8.7}_{-8.2}$&        147&        $34.4^{+5.7}_{-5.1}$&    $\leq$ 28\\
\oiv\ *&    $49.5^{+7.0}_{-6.8}$&    $94.0^{+10.4}_{-9.6}$&        --&        $28.0^{+6.3}_{-5.8}$&    --\\
\ovi\ (AOD)&    $161^{+7}_{-7}$&    $>$  $661^{+20}_{-19}$&        695&        $748^{+23}_{-22}$&    877\\
\ovi\ (PC)&    $309^{+46}_{-28}$&    $>$ $995^{+173}_{-41}$&        --&        $1480^{+680}_{-83}$&    --\\
\ovi\ (PL)&    $543^{+118}_{-83}$&    $>$ $2980^{+910}_{-140}$&    --&        $4500^{+1190}_{-140}$&    --\\
\neviii\ (AOD)&    $420^{+15}_{-14}$&    $1790^{+30}_{-30}$&        1830&        $1990^{+20}_{-20}$&    2520\\
\neviii\ (PC)&    $551^{+32}_{-26}$&    $2560^{+1480}_{-80}$&        --&        $2940^{+100}_{-80}$&    --\\
\neviii\ (PL)&    $830^{+74}_{-62}$&    $5330^{+1350}_{-110}$&        --&        $6220^{+1360}_{-140}$&    --\\
\ciii\ &    $< 6.96$&        $< 27.4$&            --&        $< 12.3$&        --\\
\niii\ &    $< 31.7$&        $< 125 $&            37&        $< 144$&        $\leq 97$\\
\oiii\ &    $< 33.6$&        $< 94.7$&            --&        $< 66.3$&        --\\
\svi\ &        $< 5.69$&        $< 13.5$&            --&        $< 27.1$&        --\\
\mgx\ &        --&            $> 1690$&            773&        $> 2150$&        1250\\
\ov\ &        --&            $> 200$&            64&        $> 170$&        141\\
 
\enddata
\label{tabmeas}
a - Column densities reported in the analysis of the absorber in \citet{Muzahid12}.
\end{deluxetable}

\clearpage

\section{Photoionization Analysis}
The ionic column densities $\left(N_{ion}\right)$ we measure are a
result of the ionization structure of the outflowing material.  These
$N_{ion}$ can be compared to photoionization models to determine the
physical characteristics of the absorbing gas.  Two main parameters 
govern the photoionization structure of each absorber: the total
hydrogen column denstity $\left(N_H\right)$ and the ionization
parameter
\begin{equation}
U_H\equiv\frac{{\displaystyle Q_H}}{{\displaystyle 4\pi R^2 \vy{n}{H} c}} \label{UEqn}
\end{equation}
(where $Q_H$ is the rate of hydrogen-ionizing photons emitted by the
object, $c$ is the speed of light, $R$ is the distance from the
central source to the absorber and $\vy{n}{H}$ is the total hydrogen
number density).  We model the photoionization structure and predict
the resulting ionic column densities by self-consistently solving the
ionization and thermal balance equations with version c08.01 of the
spectral synthesis code {\sc Cloudy}, last described in
\cite{Ferland98}.  We assume a plane-parallel geometry for a gas of
constant $\vy{n}{H}$ and initially use solar abundances and an SED
tailored for this object (other SEDs and metallicities will be
explored in Section \ref{sec:SEDZ}).  To find the pair of
$\left(U_H,N_H\right)$, defined as a phase, that best predicts the set
of observed column densities, we vary $U_H$ and $N_H$ in 0.1 dex steps
to generate a grid of models \citep[following the same approach
  described in][]{Edmonds11} and perform a minimization of the
function
\begin{equation}
\chi^2=\sum_i\left(\frac{\log (N_{i,mod}) - \log (N_{i,obs})}{\log (N_{i,obs}) - 
\log \left(N_{i,obs} \pm \sigma_i \right)} \right)^2 \label{chiSqEqn}
\end{equation}
where, for ion $i$, $N_{i,obs}$ and $N_{i,mod}$ are the observed and
modeled column densities, respectively, and $\sigma_i$ is the error in
the observed column density.

\subsection{Photoionization Modeling}\label{sec:photoModeling}

We begin by examining models of trough A that assume solar abundances and use the SED tailored to our object
(see section \ref{sec:SEDZ}). In section \ref{sec:SEDZ} we relax these assumptions and 
explore models with different metallicities and SEDs.

\begin{center}\begin{figure}[!t]
 \includegraphics[width=1.0\textwidth]{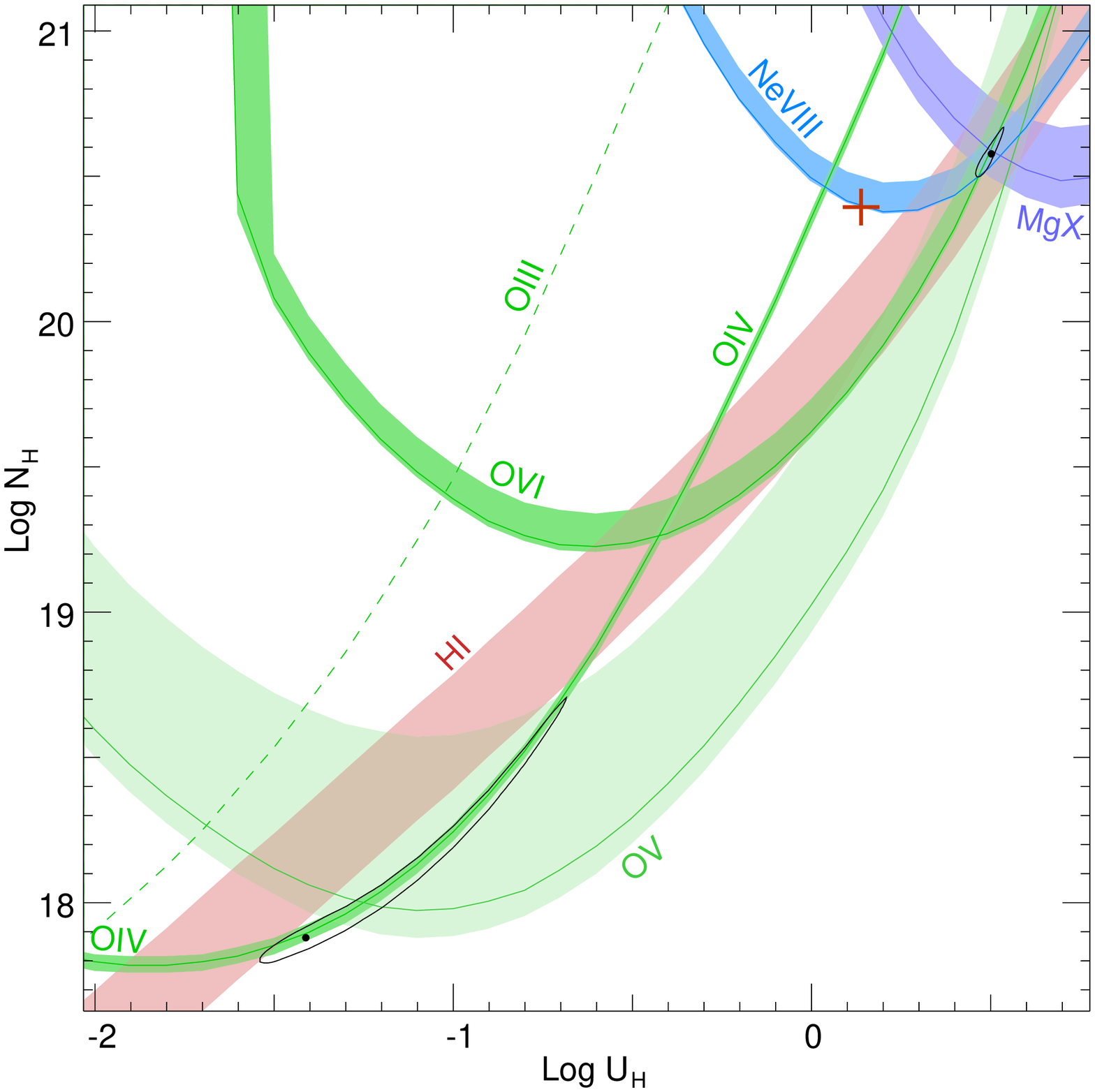}
 \caption{Phase Plot showing the ionization solution(s) for trough A
   using the HE0238 SED and assuming solar metallicity.  Each colored
   contour represents the locus of models $\left(U_{H},N_{H}\right)$
   which predict a column density consistent with the observed column
   density for that ion.  The bands which span the contours are the
   1-$\sigma$ uncertainties in the measured observations. The dashed line
   indicates the \oiii\ upper limit.  The red cross is the one-phase
   solution, while the black dots are the two-phase solutions and are
   surrounded by $\chi^2$ contours (see text).}
 \label{fig:PhasePlotA}
\end{figure}\end{center}

\begin{center}\begin{figure}[ht]
  \includegraphics[width=0.8\textwidth]{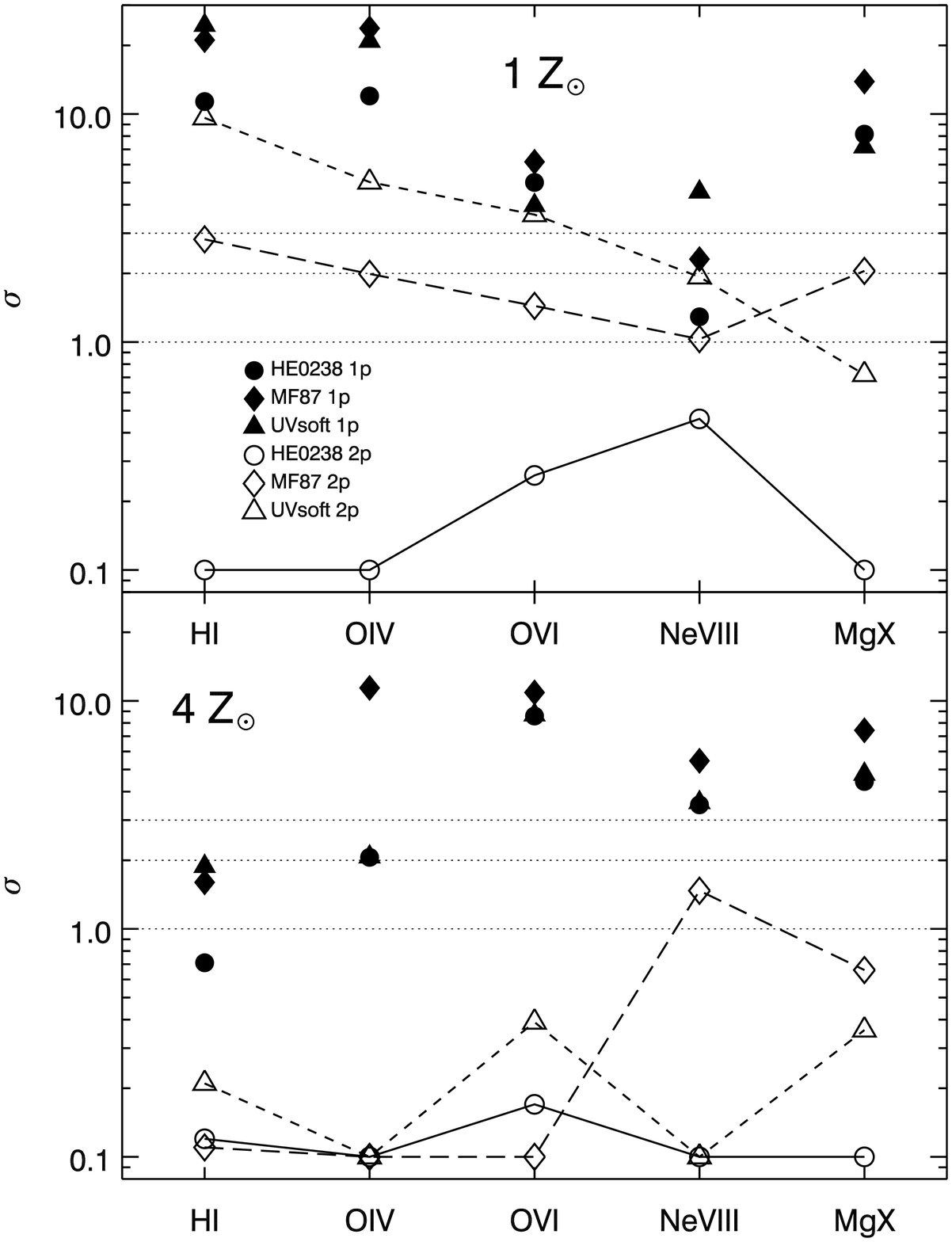}
\caption{A graphic table comparing the discrepancy between the
   modeled and observed column densities for the one-phase and two-phases
   solutions for each ion.  The top and bottom panels correspond to
   the $1Z_\odot$ and $4Z_\odot$ cases respectively, while the shape
   of the symbols in each panel represents the SED considered 
   (for SED description see Table \ref{tableLbol} and Fig.~\ref{fig:SEDs}).  The
   filled and open symbols show the error for the one- and two-phase
   solutions, respectively.  The Y-axis shows the number of standard
   deviations from the observed column densities that each model
   predicts.  The large discrepancy values of the one parameter
   solution make them physically implausible in all cases.}
 \label{fig:1vs2Parameters}
\end{figure}\end{center}

The photoionization solutions for trough A are constrained by the column
densities ($N_{ion}$) measured for \hi, \oiv, \ov, \ovi, \neviii\ and \mgx.   
where we use the $N_{ion}$ reported in Table~\ref{tabmeas}. For cases where doublet troughs are 
available we use the power-law values (PL). As we have shown in \cite{Arav08}
the PL absorber model gives a better fit in general for outflow absorption troughs than do partial covering models,
and is also more physically plausible. 
The \ciii, \niii,
\oiii\ and \svi\ upper limits are already satisfied by the upper limit
on \hi, so they do not contribute to the $\chi^2$.  A phase plot of
the results for a grid of photoionization models for trough A is
presented in Figure \ref{fig:PhasePlotA}.  The one-phase solution
(from the $\chi^2$ minimization) is marked with a red cross.
In Figure \ref{fig:1vs2Parameters} we give a graphical representation for the 
discrepancy between the predicted $N_{ion}$ of this model and our
measured ones  (filled circles in the top panel).  It is evident that the 
one-phase solution fit is unacceptable as the discrepancies for most $N_{ion}$ are between 5$\sigma$ and 10$\sigma$.
(From the other solid points on Figure \ref{fig:1vs2Parameters}, we deduce that this result is rather SED independent).

The inability of one ionization phase to provide an acceptable physical model for the 
outflow is a major conclusion of this work and therefore we discuss its robustness.
Looking at Figure \ref{fig:PhasePlotA} we observe the following: since the ratios of the oxygen ions is insensitive to
changes in chemical abundances, a one-phase ionization solution would
exist where the \oiv\ and \ovi\ curves cross in Figure
\ref{fig:PhasePlotA} (this also satisfies the \ov\ constraint as in practice,
the \ov\ $N_{ion}$ is a lower limit). However, as can be seen from the figure, such a solution underpredict
the \neviii\ and \mgx\ $N_{ion}$ by more than two orders of magnitude. In the context of a one phase model, the only way to 
arrive at an acceptable solution is to increase the relative abundances of neon and magnesium to oxygen by at least two orders of magnitude.
This, of course, is highly implausible. In the compromise $\chi^2$ one phase solution (red cross on Figure \ref{fig:PhasePlotA}) the \oiv\ $N_{ion}$ is
underpredicted by $10\sigma$ and the \ovi\ $N_{ion}$ is overpredicted
by $5\sigma$, which is unacceptable at a high confidence level

We also note that we do not use the 
\niv\ $N_{ion}$ as a constraint because its photoionization curve on Figure \ref{fig:PhasePlotA}
parallels that of \oiv\ with a 30\% lower $N_H$ value. Such a small difference can be
eliminated by a 30\%  change in the N/O abundance ratio, which is well within the allowed range in theoretical abundances models (see below). 

%\clearpage

Having demonstrated that the one-phase solution is physically
unacceptable, we next try a two-phase solution
\citep[e.g.][]{Borguet12a}.  Qualitatively, the first phase can be
found from Figure \ref{fig:PhasePlotA}, which shows that all the
contours (except for \oiv) overlap near $\log U_H\sim0.5$ and $\log
N_H\sim20.5$.  An ionization solution at that point would predict the
observed column densities for \hi, \ovi, \neviii\ and \mgx, but severely 
under-predict \oiv.  Therefore, a second, low-ionization phase is
required to predict the observed \oiv\ without adding significantly to
the predictions of the other ions.  Such a solution would appear along
the \oiv\ contour and below the \ov\ contour in Figure
\ref{fig:PhasePlotA}.  The actual two-phase solution can be found by
minimizing the $\chi^2$ function given by equation (\ref{chiSqEqn}),
where $N_{i,mod}$ becomes the sum of the predicted column densities
from both ionization solutions.  The best two-phase solution is shown
in Figure \ref{fig:PhasePlotA} as black dots at $\log
U_{H,low}=-1.4^{+0.7}_{-0.1}$, $\log N_{H,low}=17.9^{+0.8}_{-0.1}$,
$\log U_{H,high}=0.50^{+0.04}_{-0.04}$ and $\log
N_{H,high}=20.6^{+0.1}_{-0.1}$.  The black dots are surrounded by
contours representing the solutions with $\chi^2=\chi^2_{min}+1$.  If
$\chi^2_{min}\ge 1$, then the contours represent solutions with
$\chi^2=2\chi^2_{min}$, which is equivalent to renormalizing to
$\chi^2_{min}=1$ and adding $\Delta\chi^2=1$.  Note that the errors
are correlated as is demonstrated graphically in Figure
\ref{fig:PhasePlotA}.  The column density discrepancy for the
two-phase solution is illustrated in the top panel of Figure
\ref{fig:1vs2Parameters} with open circles, and is below $1\sigma$ for
each ion.  It is clear that a two-phase solution fits the observations
better than a one-phase solution and is an acceptable physical model.
This result also holds for the model variations discussed in the next
section.

\subsection{Solution Dependency on Spectral Energy Distribution and Metallicity}\label{sec:SEDZ}
\begin{center}\begin{figure}[ht]
 \includegraphics[angle=-90,width=1.0\textwidth]{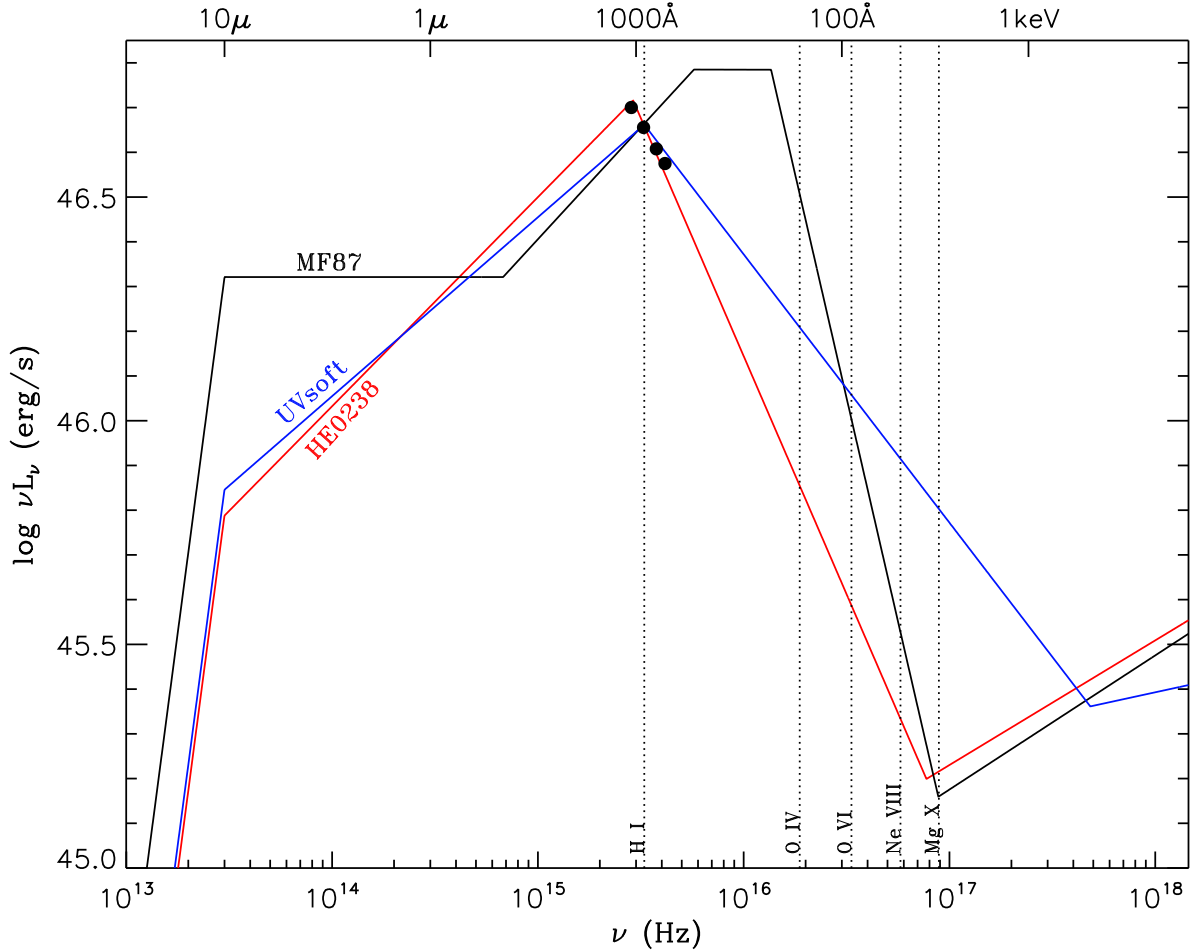}
\caption{Comparison of the SEDs used in the analysis of HE0238-1904.  
The black dots represent the COS data used to scale the SEDs, and the dotted 
lines show the ionization destruction potential for some of the ions used to 
constrain the photoionization solution.  The HE0238 SED is used as the ``standard'' SED for our models.}
 \label{fig:SEDs}
\end{figure}\end{center}

The photoionization and thermal structure of an outflow depends on the
spectral energy distribution (SED) incident upon the outflow.
Traditionally the MF87 SED \citep{Mathews87} is used to model
radio-loud quasars since, as \cite{Dunn10a} explains, the samples of
``typical'' quasars used to construct the MF87 SED were dominated by
radio-loud QSOs.  \cite{Telfer02} found, from a sample of radio-quiet
quasars, a much softer continuum compared to the MF87 SED.  Since
HE0238-1904 is a radio-quiet quasar, the slope of the observed
continuum indeed suggests a lack of the so-called big blue bump
characteristic of the MF87 SED (the solid black line in Figure
\ref{fig:SEDs}).  For this reason, we developed an SED appropriate for
HE0238-1904 by matching the far-UV shape to the COS data and largely
mimicking the standard MF87 SED at other energies.  As an added
comparison, we include the UV-soft SED previously used for high
luminosity, radio-quiet quasars described in \cite{Dunn10a}.  These
three SEDs are shown in Figure \ref{fig:SEDs}, and their bolometric
luminosity $\left(L_{bol}\right)$ and emission rate of hydrogen
ionizing photons $\left(Q_H\right)$ are listed in Table
\ref{tableLbol}.  Calculations were made using the fitted continuum
model for the observed flux of the object (see Section 
\ref{sec:emissionModel}) which gives $F_{1230}=1.57\times 10^{-14}
\mathrm{\ ergs\ s}^{-1}\mathrm{\ cm}^{-2}\ \mathrm{\AA}^{-1}$ and
using a standard cosmology ($H_0=73.0
\mathrm{\ km\ s}^{-1}\mathrm{\ Mpc}^{-1}$, $\Omega_\Lambda=0.73$, and
$\Omega_m=0.27$).

\begin{deluxetable}{lrrr}
\tablecaption{\sc Properties of the chosen SEDs}
\tablewidth{0pt}
\tablehead{\colhead{SED}&\colhead{HE0238}&\colhead{MF87}&\colhead{UVsoft}}
\startdata
$L_{bol}\ \left(10^{47}\ \mathrm{erg\ s}^{-1}\right)$&$1.50$&$1.62$&$1.39$\\
$Q_H\ \left(10^{57}\ \mathrm{s}^{-1}\right)$&$0.841$&$1.33$&$0.992$\\
\enddata
\label{tableLbol}
\end{deluxetable}

\clearpage

\begin{figure}[ht]
  \includegraphics[width=0.8\textwidth]{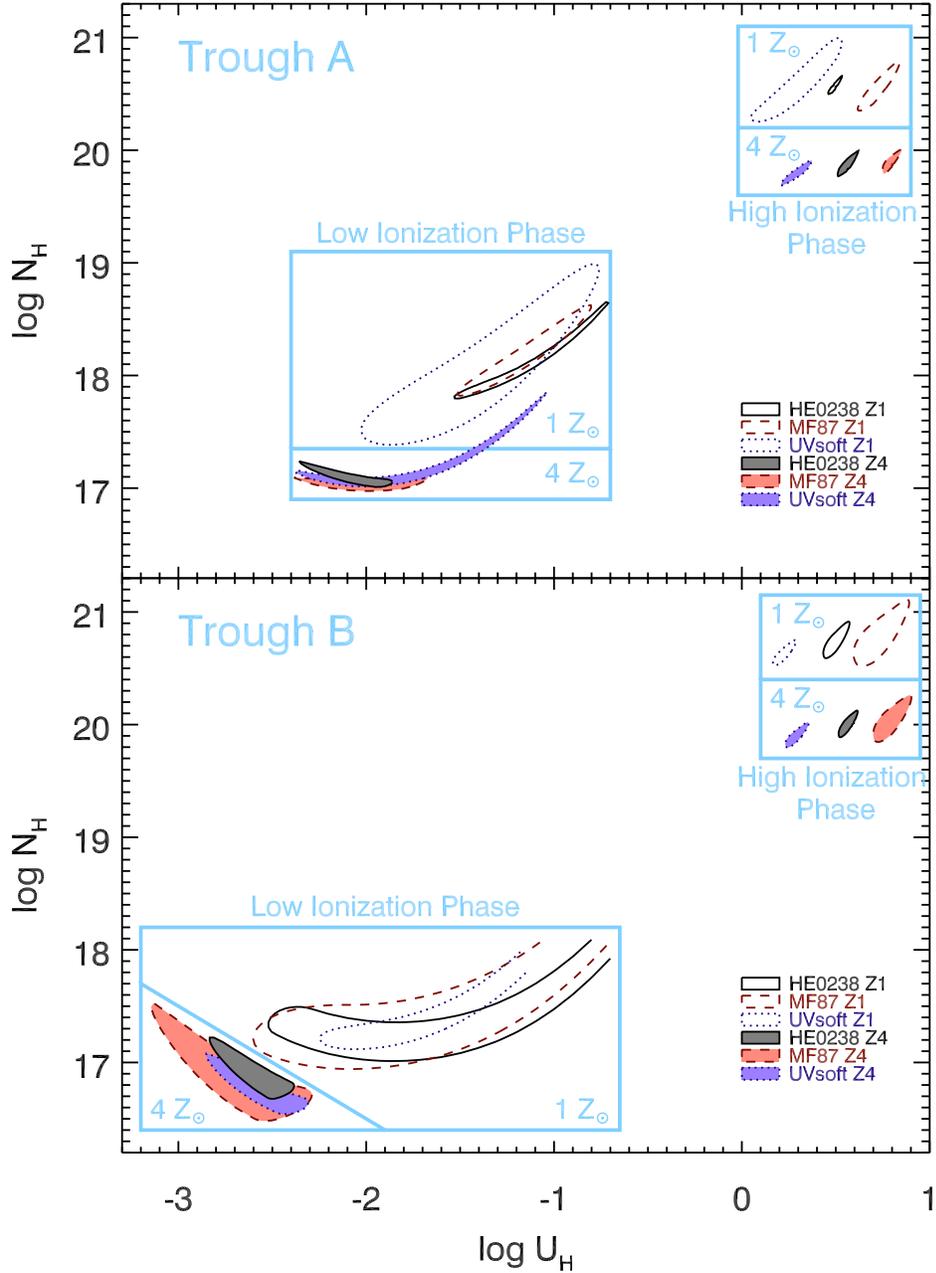}
\caption{Phase Plot for the two-phase ionization solutions for the 2
   troughs, 3 SEDs and 2 metallicities for a total of 12 models.  All
   the models agree on the existence of a two-phase outflow.  The
   $\chi^2$ contour for the low-ionization phase of the $1Z_\odot$
   models for trough B can continue up to $\log N_H\sim19.7$ and $\log
   U_H\sim-0.4$ (see text).  Increasing the metallicity from
   $1Z_\odot$ to $4Z_\odot$ lowers the hydrogen column density of both
   components roughly inversely linearly.  The SED shape only affects
   the high-ionization component by shifting its ionization parameter
   by $\sim0.2$ dex in either direction compared to the standard
   HE0238 SED.}
 \label{fig:PhasePlotAB}
\end{figure}

We consider the HE0238 SED the most physically plausible SED for this
object, while the MF87 or UVsoft SEDs are used to demonstrate the
sensitivity of the photoionization solution to other QSO SEDs used in
the literature.  The effect of the different SEDs can be seen in the
high-ionization phase in Figure \ref{fig:PhasePlotAB}, where the MF87
and UVsoft SED increases and decreases the ionization parameter
relative to the HE0238 SED by $\sim0.2$ dex, respectively.  These
differences arise since the high-ionization phase is dominated by the
\ovi, \neviii\ and \mgx, whose ionization destruction potentials occur
in a frequency range (see Figure \ref{fig:SEDs}) where the three SEDs
differ in power-law indices.

The other main parameter that affects the photoionization and thermal
structure of an outflow is the metallicity of the gas.  AGN outflows
are known to have supersolar metallicities (e.g. Mrk 279:
$Z\simeq2Z_\odot$, \citen{Arav07}; SDSS J1512+1119: $Z_\odot\ltorder
Z\ltorder4Z_\odot$, \citen{Borguet12b}; SDSS J1106+1939: $Z=4Z_\odot$,
\citen{Borguet13}).  We therefore consider the effects of different
metallicities on the ionization solution
for each SED, using the elemental abundance scaling defined for {\sc
  Cloudy} starburst models \citep[which follows grid M5a
  of][]{Hamann93}.  Increase in metallicity (see Figure
\ref{fig:PhasePlotAB}) lowers the total hydrogen column density
$\left(N_H\right)$ for both ionization phases roughly inversely
linearly.  This effect can be used to constrain the metallicity of the
outflow by lowering the model's metallicity until the \neviii\ contour
has shifted above the \hi\ band in Figure \ref{fig:PhasePlotA}.  This occurs at $Z\simeq\frac{1}{4}Z_\odot$ and 
does not allow the high-ionization phase model to fit both the \hi\ and
\neviii\ measurements, which result in a large $\chi^2$ and make the
solution physically implausible.  

The upper limit on the metallicity
can be found by raising the metallicity until both ionization phases
cannot produce the \hi\ $N_{ion}$ lower limit.  At this metallicity, the high-ionization
phase would still have to be at the crossing of \mgx, \neviii\ and \ovi.
The low-ionization solution must be along the \oiv\ contour and below the
 \oiii\ upper limit (If the solution is above the \oiii\ upper limit we should detect an \oiii\ trough.).
Therefore, the low-ionization solution the maximizes the \hi\ $N_{ion}$ and does not 
violate other $N_{ion}$ constraints, is found at the crossing point of the 
 \oiv\  measurement and the \oiii\ upper limit contours. We find that this transition occurs at 
$Z\simeq 4Z_\odot$. At higher $Z$ values, the discrepency between the measured \hi\ $N_{ion}$ 
and the total \hi\ $N_{ion}$ from both phases, leads to unaccepted lagre values of $\chi^2$.
Combing the upper an lower limits on the metallicity, we obtain a possible range of 
$\frac{1}{4}Z_\odot\ltorder Z\ltorder 4Z_\odot$. 

\clearpage

%\begin{center}
%\end{center}

\clearpage

Trough B at $\sim -5000$ \kms\ is  kinematically separated  from trough A ($\sim -3850$ \kms), as both have 
$\Delta v\sim500$ \kms\ (see Figure \ref{all_line}). However, we find that these two troughs have strong  
resemblence in their physical characteristics.  Both have similar photoionization solutions, as is evident by the 
two panels of Figure \ref{fig:PhasePlotAB}. For the SEDs and metallicities
considered, both troughs show clear evidence for the existence of a two ionization-phase
outflow separated by $\Delta U_H\sim2 \mathrm{dex}$ and $\Delta
N_H\sim2.5 \mathrm{dex}$.  They also have 
indistinguishable number density for their low ionization phase (see next section).

For the solar metalicity case, the low ionization solution for trough B can be anywhwere along the 
\oiv\ contour up to $\log N_H\sim19.7$ and $\log U_H\sim0.0$ 
(extending the $\chi^2$ contour shown in Figure \ref{fig:PhasePlotAB}, for the $1Z_\odot$ cases).
The reason for this behavior is that for $1Z_\odot$ all the other $N_{ion}$ are well
predicted by the high-ionization phase, including \hi. Therefore, any position along the \oiv\ contour,
which is well below the \hi\ band will yield the desired $N$(\oiv), while contributing negligibly to the other 
predicted $N_{ion}$.  For higher metalicity, the  high-ionization phase cannot produce enough $N$(\hi), 
therefore most of the \hi\ must come from the low ionization phase, 
which cause it to place the solution close to the crossing point of the 
\oiv\ contour and \hi\ band.  We note that in all these cases a lower ionization phase is necessary to
produce not only the observed \oiv\, but also the observed \niv.
Using the same methods from trough A, we obtain a possible range of 
$\frac{1}{4}Z_\odot\ltorder Z\ltorder 3Z_\odot$. 

\subsection{Comparison With the Results of \citet{Muzahid12}}

\citet[hereafter M2012]{Muzahid12} analyzed the same observations of HE0238-1904.
We compare their ionic column density ($N_{ion}$) measurements to ours in Table \ref{tabmeas}.
Most of the measurements agree to better than a factor of two 
(particular good agreement for the AOD measurements of the deepest COS troughs: \ovi\ and \neviii).
However, as expected, our power-law measurements are consistently larger by roughly a factor of 3.

M2012 found that the outflowing material cannot be in one ionization phase; a major finding that we concur with.
A quantitative comparison between our results and those reported by M2012 is complicated due to the different approaches 
both groups have taken in the analysis. M2012 photoionization analysis concentrated on finding $U_H$ values that 
reproduced individual ratios of $N_{ion}$ measurements for each of their 5 labeled components (see their Figure 4 and accompanying text). 
Our approach, as detailed above, is to find a full two-phase photoionization model that simultaneously accounts for all the measured $N_{ion}$, 
in each of the troughs A and B. With these differences in mind, we note that for the high ionization phase, M2012 obtains $0.8<\log U_H<1.2$ for components
3, 4 and 5 (covering the bulk of both A and B troughs), less than a factor of 3 difference from our results of $0.2<\log U_H<0.8$ (see Fig. \ref{fig:PhasePlotAB}).
The low ionization phase is less constrained in M2012, as they only use comparison of high ionization to low ionization $N_{ion}$ 
(i.e., \ovi/\oiv\ and \ovi/\niv). It is thus not surprising that our $U_H$ for the low ionization phase are roughly two orders of 
magnitude lower than the constraints achieved from the \ovi/\oiv\ and \ovi/\niv\ ratios. We cannot compare $N_H$ values as they are not reported by M2012.

\section{ENERGETICS}

We use equation set \eqref{energetics} to estimate the mass flow rate and kinetic luminosity of the outflow.
As described in section 2, to do so we need to measure or constrain the values for $N_H$, $R$ and $\Omega$.
The first of these  ($N_H$) was determined by fitting the $N_{ion}$ measured in section \ref{coldens_absmo} 
with the prediction from a  grid of {\sc Cloudy} simulations (see previous section). 
The ionization parameter $U_H$ was determined simultaneously with
$N_H$, and will be used in section 7.1 to determine the distance between
the absorber and the central source.  Constraints on global covering fraction $\Omega$ 
are elaborated on in section 7.2. Following these preliminary steps we derive and discuss the energetics of the outflow components
(Section 7.3 and Table \ref{tab:Energetics})

\subsection{Determining the Number Density and
Distance of the Outflow}\label{sec:numDenDist}

\begin{center}\begin{figure}[ht]
 \includegraphics[width=0.8\textwidth]{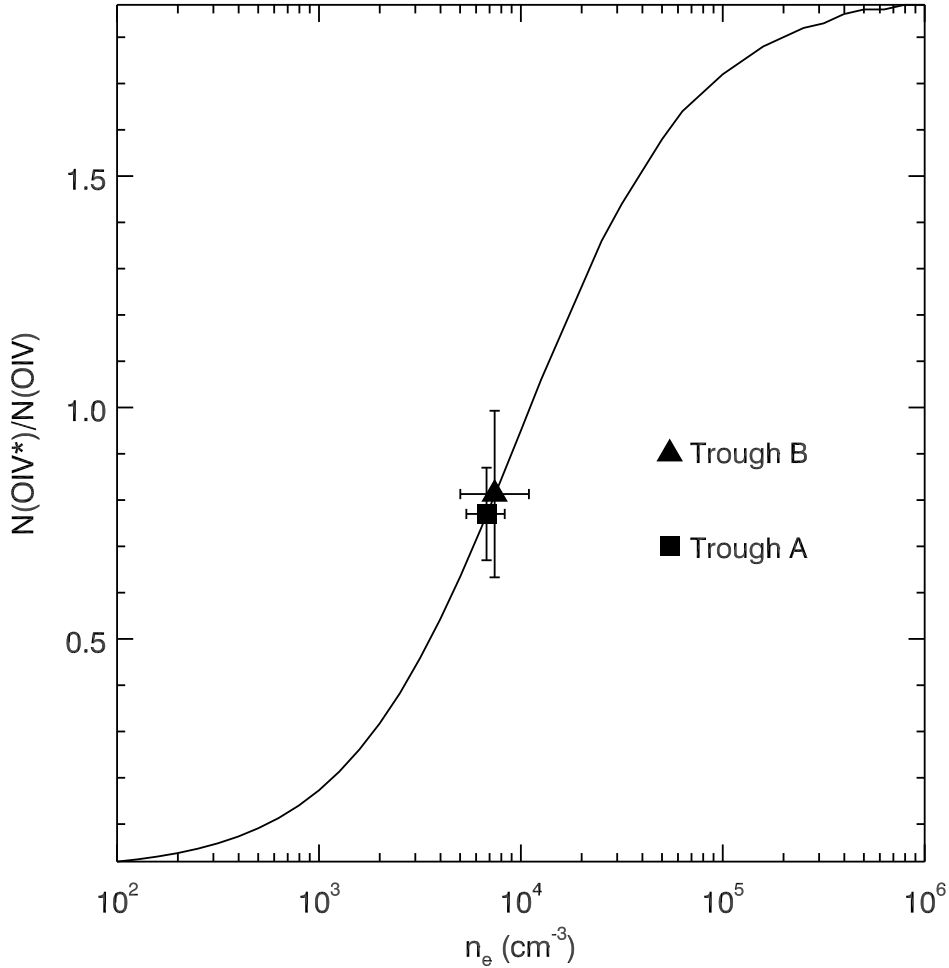}
\caption{Density diagnostic for HE0238-1904 trough A and B.  We overlay their
measured $N_{ion}$(\oiv*)/$N_{ion}$(\oiv) on the theoretical level
population assuming a temperature of $10^4$K, to obtain the electron
number densities $\left(n_e\right)$ inferred from the collisional excitation
model.  The $n_e$ is translated to $\vy{n}{H}$ via the relation $n_e \simeq 1.2
\vy{n}{H}$, valid for highly ionized plasma.}
 \label{fig:ExcitedOIV}
\end{figure}\end{center}

In highly ionized plasma the hydrogen number density is related to the electron
number density through $n_e \simeq 1.2 \vy{n}{H}$.  To determine $n_e$ for trough A we use the measured the $N_{ion}$(\oiv*)/$N_{ion}$(\oiv) ratio
in trough A3 (the portion of
trough A where the excited state appears) and similarly for trough B.
Figure \ref{fig:ExcitedOIV} shows the observed ratios between \oiv* and
\oiv\ plotted on the theoretical dependence of this ratio on $n_e$, for both troughs.  The error on the observed column
densities translates into errors on the $n_e$, yielding a measurement of
$\log\left(\vy{n}{H}\right)=3.75\pm 0.10 \mathrm{\ cm}^{-3}$ for trough A and
$\log\left(\vy{n}{H}\right)=3.79\pm 0.17 \mathrm{\ cm}^{-3}$ for trough B. We note that both troughs 
have the same number density within the measurement error.

The $N_{ion}$(\oiv*)/$N_{ion}$(\oiv)  ratio was derived using a sub portion of each trough.
Therefore, we need to address the physical relationship between these sub-components and the more extended troughs. 
The kinematic correspondence, suggests a co-spatial location for the gas that shows 
\oiv* and \oiv\ with the full parent trough. We can further solidify this connection by 
comparing the overall photoionization solution of each subtrough to its parent trough.
We find that there is a very good agreement in both cases.  For example:
The photoionization solution for trough A3 is virtually identical 
to that of trough A, as the ionization parameters for both the high  and
low ionization phases differ by only $\sim0.1$ dex.  This strong physical resemblance combined with 
the kinematic correspondence solidifies the case that trough A3 is indeed a physical portion of outflow component A.
A similar situation exists for trough B.

We determine $R$ for each component by substituting the measured $\vy{n}{H}$ 
combined with the $U_H$ of each low ionization phase (where all the \oiv\ gas resides), 
and the $Q_H$ given in table \ref{tableLbol} into equation \eqref{UEqn}.
The results for all cases are shown in Table \ref{tab:Energetics}.

\subsection{Constraining $\Omega$}
\label{omegadisku}

For a full discussion about constraining $\Omega$ that is needed for
equation set (1), we refer the reader to Section 5.2
in \citet{Dunn10a}. Here we reproduce the main arguments and then
concentrate on the case of \neviii\ outflows. There is no direct way to
obtain the $\Omega$ of a given outflow from
its spectrum as we only see the material located along the line of
sight. Therefore, the common procedure is to use the fraction
of quasars that show outflows as a proxy for  $\Omega$. 

A significant amount of work was directed towards finding the frequency of detecting different 
categories of quasar outflows from the full population of quasars \citep{Hewett03,Ganguly08,Dai08,Knigge08,Dunn12,Dai12}.
However, none of these studies addressed the frequency of quasar outflows that show troughs from \neviii, 
the highest abundant ionization species seen in the UV.  Recently, this issue was addressed by \citet{Muzahid13}
who found 11 \neviii\ outflows in 20 quasar spectra. The width of these outflow troughs is mainly between 200--700 \kms, 
straddling the $\sim450$ \kms\ width we measure for troughs A and B in HE~$0238-1904$.  We therefore use $\Omega=0.5$
as a representative for this class of outflows.

\subsection{Results}\label{sec:Results}

For the analysis of the 2 observed troughs, we have considered 3 different SEDs
combined with 2 possible metallicities for a total of 12 model variations. 
In Table \ref{tab:Energetics} we detail the main results for each model.
The 1st column designates the SED used; the 2nd column gives the $\log\left(U_H\right)$
of the low-ionization phase, since this is the $U_H$ that determines the distance of the absorber (see section \ref{sec:numDenDist}); 
column 3 gives the $\log\left(N_H\right)$ of the high-ionization phase, 
as it carries the vast majority of outflowing column density; in the 4th column we provide the distance of the 
outflow component using the $U_H$ reported in the 2nd column and the calculation detailed in section \ref{sec:numDenDist};
the 5th through 7th columns give the mass flow rate, kinetic luminosity and its percentage from the total $L_{\mathrm{Bol}}$ of the quasar, by substituting 
$N_H$ and $R$ from columns 3 and 4 into equation set (1), and using $\Omega=0.5$ (see section \ref{omegadisku})

The results of these models can be summarized as follows: \\
{\it 1. Trough A vs Trough B:}  trough B carries roughly 5 times more
kinetic luminosity than trough A.  This is due to its $30\%$
greater velocity, $70\%$ higher $R$ and a $25\%$ larger $N_H$. \\
{\it 2. Solar vs Supersolar Metallicity:}  the higher metallicity models result in
a lower $N_H$ by a factor roughly inversely linear with the metallicity.  For trough A, the ionization parameter
for the low-phase also decreases, suggesting that the outflow could be twice as
far from the central source if a higher metallicity is assumed.  These combined
effects result in a kinetic luminosity 3 times lower than that of solar
metallicity models. \\
{\it 3. SED dependence:}  The energetics are rather insensitive to the SED shape.
 The ionization parameter of the high-phase changes by up to a factor of 4, but the $N_H$ is unaffected.
For the low-ionization phase the differences in $U_H$ are all within the allowed $\chi^2$ contours: no apparent trend is detected.

We emphasize that the HE0238 SED we use is the most appropriate given the spectral data on hand. In addition,
the $Z=4Z_\odot$ case gives a slightly better fit to the
measured column densities than the pure solar case.  At the same time, it
provides a more conservative (lower) estimate for the kinetic luminosity of this
outflow.  We therefore use this model as the representative result for the
outflow.

For the two separate outflow components of HE0238-1904 (troughs A and B), we
select trough B as the representative of the energetics contained within this
object, since trough B has roughly 5 times more energy than trough A.  We show
this result in Table \ref{tabmflo} along with the results from our previous
$\dot{E}_k$ record holders:  SDSS J1106+1939 \citep{Borguet13} and SDSS
J0838+2955 \citep[][but see the correction by a factor of 0.5 reported in
\citen{Edmonds11}]{Moe09}.

 \begin{deluxetable}{rrrrrrr}
\tablecaption{\sc Model comparison}
\tablewidth{0pt}
\tablehead{
\colhead{SED}&\colhead{$\log\left(U_H\right)$}&\colhead{$\log\left(N_H\right)$}&\colhead{$R$}&\colhead{$\dot{M}$}&\colhead{$\dot{E}_k$}&\colhead{$\dot{E}_k$/$L_{\mathrm{Bol}}$}\\
\colhead{}&\colhead{}&\colhead{($\mathrm{cm}^{-2}$)}&\colhead{(kpc)}&\colhead{$\left(\mathrm{M}_\odot \mathrm{yr}^{-1}\right)$}&\colhead{$\left(10^{45} \mathrm{\ erg\ s}^{-1}\right)$}&\colhead{(\%)}
}
\startdata
&&&\multicolumn{4}{l}{\ \ {\bf Trough A} \hspace{0.35in}($v=-3850\ \mathrm{km\ s}^{-1}$)}\\
\hline
\multicolumn{7}{c}{$\mathbf{Z=1Z_\odot}$}\\
HE0238&$-1.4^{+0.7}_{-0.1}$&$20.6^{+0.09}_{-0.09}$&$1^{+0.29}_{-0.57}$&$27^{+10}_{-15}$&$0.522^{+0.2}_{-0.3}$&$0.35^{+0.1}_{-0.2}$\\
MF87&$-1.1^{+0.3}_{-0.4}$&$20.5^{+0.2}_{-0.2}$&$0.91^{+0.61}_{-0.28}$&$22^{+25}_{-10}$&$0.43^{+0.5}_{-0.2}$&$0.27^{+0.3}_{-0.1}$\\
UVsoft&$-1.3^{+0.5}_{-0.7}$&$20.5^{+0.5}_{-0.3}$&$0.99^{+1.3}_{-0.47}$&$23^{+70}_{-14}$&$0.452^{+1}_{-0.3}$&$0.32^{+0.9}_{-0.2}$\\
\multicolumn{7}{c}{$\mathbf{Z=4Z_\odot}$}\\
HE0238&$-2^{+0.1}_{-0.4}$&$19.8^{+0.2}_{-0.06}$&$2^{+1.2}_{-0.32}$&$9.3^{+8}_{-2}$&$0.179^{+0.1}_{-0.04}$&$0.12^{+0.1}_{-0.02}$\\
MF87&$-1.9^{+0.2}_{-0.5}$&$19.9^{+0.1}_{-0.09}$&$2.2^{+1.9}_{-0.49}$&$11^{+11}_{-3}$&$0.22^{+0.2}_{-0.06}$&$0.14^{+0.1}_{-0.04}$\\
UVsoft&$-1.8^{+0.8}_{-0.6}$&$19.8^{+0.2}_{-0.07}$&$1.8^{+1.8}_{-1}$&$6.9^{+8}_{-4}$&$0.133^{+0.2}_{-0.08}$&$0.096^{+0.1}_{-0.06}$\\
\hline
&&&&&&\\
&&&\multicolumn{4}{l}{\ \ {\bf Trough B} \hspace{0.35in}($v=-5000\ \mathrm{km\ s}^{-1}$)}\\
\hline
\multicolumn{7}{c}{$\mathbf{Z=1Z_\odot}$}\\
HE0238&$-2.4^{+2}_{-0.1}$&$20.8^{+0.1}_{-0.2}$&$3^{+0.88}_{-2.8}$&$160^{+80}_{-150}$&$5.32^{+3}_{-5}$&$3.5^{+1.7}_{-3}$\\
MF87&$-2.2^{+2}_{-0.3}$&$20.9^{+0.2}_{-0.4}$&$3.1^{+1.7}_{-2.8}$&$210^{+200}_{-200}$&$6.86^{+7}_{-6}$&$4.2^{+4}_{-4}$\\
UVsoft&$-1.8^{+1}_{-0.4}$&$20.7^{+0.09}_{-0.1}$&$1.7^{+1.2}_{-1.2}$&$69^{+50}_{-50}$&$2.25^{+1.8}_{-1.7}$&$1.6^{+1.3}_{-1.2}$\\
\multicolumn{7}{c}{$\mathbf{Z=4Z_\odot}$}\\
HE0238&$-2.5^{+0.09}_{-0.4}$&$20^{+0.1}_{-0.1}$&$3.4^{+2}_{-0.49}$&$31^{+20}_{-9}$&$1^{+0.7}_{-0.3}$&$0.67^{+0.5}_{-0.2}$\\
MF87&$-2.5^{+0.2}_{-0.7}$&$20^{+0.2}_{-0.2}$&$4.2^{+5.2}_{-0.86}$&$40^{+60}_{-15}$&$1.29^{+2}_{-0.5}$&$0.8^{+1.3}_{-0.3}$\\
UVsoft&$-2.4^{+0.1}_{-0.4}$&$19.9^{+0.1}_{-0.1}$&$3.4^{+2.5}_{-0.51}$&$24^{+20}_{-6}$&$0.772^{+0.7}_{-0.2}$&$0.56^{+0.5}_{-0.13}$\\
\hline
\enddata
\label{tab:Energetics}
\end{deluxetable}

\nopagebreak

\section{A TWO PHASE OUTFLOW}\label{sec:twophase}

One of the important findings of this paper is the existence of at least two
widely different ionization phases in the outflow.
For both outflow components we find that the two phases differ by roughly 2.5--3
orders of magnitude in $U_H$ and similarly in $N_H$ 
(where the higher ionization phase has the larger $N_H$). The robustness of this
result relies on: a) the inability of a single phase solution 
to produce a reasonable fit for the measured $N_{ion}$, b) an excellent fit
achieved by the two-phase models, c) reliable $N_{ion}$ extractions, and 
d) a need for a two-phase solution that is independent of the assumed SED or
metalicity. In addition, the high spectral resolution of the data 
plus the ionization modeling of the different sub-components give high
confidence to the assertion that the two phases are co-spatial in both
components.

\subsection{Geometry and Filling Factor}\label{sec:geometry}

An interesting physical picture of the outflow emerges from the two-phase
solution. For component B (using our standard HE0238 SED and $4Z_\odot$ metalicity) we
find that the high-ionization phase (hereafter phase 2) carries a thousand times more column density
than the low-ionization phase (hereafter phase 1). At the same time
 the two phases differ in $U_H$ by a similar factor. Therefore, 
phase 1 is inferred to have a volume filling factor of $\sim10^{-6}$
embedded within a
much more massive and extended phase 2. We also determined that in
the hot phase $\log\left(\vy{n}{H}\right)=0.8 \mathrm{\ cm}^{-3}$  
(from our determination of $\vy{n}{H}$ for the phase 1 and their
difference in $U_H$) and $N_H\sim10^{20}\mathrm{\ cm}^{-2}$.
Therefore, the thickness of this outflow is $\Delta R=N_H/\vy{n}{H}\simeq5pc$. 
As we show in section \ref{sec:comparison}, x-ray observations of the NGC 3783 outflow 
show an even higher ionization phase to which the UV
data of HE0238-1904 is not sensitive. If a similar phase exists in the HE0238-1904 outflows then the 
full width of the outflow will be $\Delta R\simeq70pc$ (see section \ref{sec:comparison}).

\subsection{Are the Two Phases in Pressure Equilibrium?}\label{sec:pressure}

For component A (using our standard HE0238 SED and $4Z_\odot$ metalicity)
our {\sc Cloudy} models, which assume only photoionization equilibrium, give a temperature of
$7\times10^3K$ and $7\times10^4K$ 
for phase 1 and 2 , respectively. Such a situation is far
from pressure equilibrium between the two 
phases, since the factor 10 higher temperature cannot compensate for the factor
of 300 lower density of phase 2 compared to phase 1
(inferred from the difference in their $U_H$). Therefore, 
if the ionization equilibrium and thermal balance of the outflow are caused only
by photoionization, 
the two phases cannot be in pressure equilibrium. This will make the
phase 1 material unstable on roughly its dynamical time-scale for
expansion:
$t=D/v_{th}\ltorder [N_H/\vy{n}{H}]/v_{th}\simeq [10^{17}/10^4]/10^6\simeq10^7$
seconds, which is less than a year 
(where $D$,  $N_H$, $\vy{n}{H}$ and $v_{th}$ are the maximum size, column
density, number density and thermal velocity of the low-ionization absorber,
respectively).

However, there are at least 3 mechanisms suggested in the literature that 
can potentially put the two phases in pressure equilibrium: a) magnetic
confinement of the outflow \citep{dekool95}
b) cloudlets compressed by hot post-shock gas \citep{Faucher-Giguere12}; and c)
thermal pressure equilibrium with the high ionization phase, whose
temperature is much higher than the photoionization equilibrium indicates
(presumably from shock heating).  

We did some preliminary tests of the last
hypothesis (c) and found the following.  Having the two phases in pressure
equilibrium 
is equivalent to the requirement that $T_2=T_1(U_{H2}/U_{H1})$ where the
subscript 1 and 2 
refers to the low-ionization and high-ionization-phase, respectively. 
{\sc Cloudy} models give us $T_1$, and the number density of this phase ($n_1$)
is known from  the $N_{ion}$(\oiv*)/$N_{ion}$(\oiv) ratio.
From the definition of $U_{H}$ we then obtain $n_2$=$n_1(U_{H1}/U_{H2}$).
We therefore proceed by running a {\sc Cloudy} model where $U_{H}=U_{H2}$,
$\vy{n}{H}=n_2$ while forcing $T=T_2$.
For component A we find that collisional ionization at $T_2$ (a few $10^6$ K)
causes the ionic fraction of \ovi, \neviii\ and \mgx\ to drop by  
3.3, 2.8 and 1.8 dex, respectively, compared to the pure photoionization
equilibrium of phase 2.  Such reduction in ionic populations with the same $N_H$ will not allow for 
enough $N_{ion}$ to detect any troughs from these species. Simply increasing the $N_H$ of 
phase 2 by two or three orders of magnitude would also not work as we either under-predict 
\ovi\ or over-predict \neviii\ and \mgx.

One possible way for this mechanism to work and still yield the required $N_{ion}$
for phase 2 is as follows. Figure \ref{fig:PhasePlotA} shows that an acceptable solution for 
phase 1 can be found along the \ovi\ contour, up to $\log\left(U_H\right)=-0.7$.
If we move the solution farther to the crossing point of the \oiv\ and \ovi\ contours 
[$\log\left(U_H\right)=-0.4, \log\left(N_H\right)=19.2$], phase 1  produces all the observed 
\ovi\ $N_{ion}$ by construction. We now allow phase 2 to have a temperature high enough 
to be in pressure equilibrium with phase 1 ($T_2=2.8\times10^5$K, calculated from the $\Delta U_H$ between the phases), 
and increase its total $N_H$ by 0.4 dex.
For this model we find that the $N_{ion}$ for \neviii\ matches the observations exactly and that of \mgx\ is 
0.5 dex above the measurements, which is an acceptable solution (see table \ref{tabmeas}). At this temperature and density, 
collisional ionization decreased the fraction of \neviii\ by 0.4 dex and increased that of \mgx\ by 0.1 dex.
The fraction of \ovi\ in phase 2 dropped by 1.0 dex, however, and we now obtain all the observed \ovi\ $N_{ion}$
from phase 1.  To summarize, using the following assumptions: a heating mechanism that increases the temperature of 
phase 2 by roughly a factor of 3 above what a pure ionization equilibrium yields, increases the total $N_H$ of phase 2 by 0.4 dex,
and moves the solution of phase 1 to the crossing point of the \ovi\ and \ovi\ contours; 
we can obtain an acceptable solution for all the observed $N_{ion}$ while keeping both phases in pressure equilibrium.
Such a solution will put the outflow closer to the central source by a factor of three (from the difference in $U_{H1}$), 
but will not change the overall mass flux and kinetic luminosity significantly since the  $N_H$ of phase 2 will increase by a similar factor. 
However, the filling factor for phase 1 (see section \ref{sec:geometry}) will increase from  $\sim10^{-5}$ to $\sim10^{-3}$.

\subsection{Comparison With Other AGN Outflows}\label{sec:comparison}

The ionization species which yield outflow troughs in  
most UV AGN spectra cover a narrow range in ionization potential (IP), 
usually from \siiv\ (45 eV) to \nv\  (98 eV) and sometimes \ovi\
(138 eV). This small range allows for a good fit of the measurements with a single
ionization phase \citep[e.g.,][]{Edmonds11}. With coverage of the far UV we can observe \neviii\ (239 eV)
and \mgx\ (368 eV). The larger range in IP can indicate that at least a two-phase medium is needed. 
We find only one case in the literature where data of \neviii\ was shown to
suggest a two-phase solution \citep[PG 0946+301,][]{Arav01}, with the low-ionization
phase at $-2<\log\left(U_H\right)<-1.5$ and the high ionization phase at 
$\log\left(U_H\right)\sim0.5$ (deduced from their figure 7). These values are similar to those we find in
HE0238-1904.

X-ray observations
of Seyfert outflows, the so-called ``warm absorbers'', routinely show the need
for two or more ionization phases \citep[e.g.,][]{Netzer03, Steenbrugge05,Holczer07}. 
This is mainly due to the much larger spread in IP of the observed ionic
species [(e.g., \ov\ (IP=114 eV) to \oviii\ (IP=871 eV)]. However, in almost all cases,
the x-ray spectra lack the resolution to kinematically associate the warm
absorber with the UV absorber seen in the same object.
Two comprehensive analyses of the UV and X-ray photoionization equilibrium of
outflows from the same object exist in the literature. For Mrk 279, \citet{Costantini07} and
\citet{Arav07} show a remarkable agreement for the properties of the low-ionization
phase of the outflow: $\log\left(U_H\right)=-1, \log\left(N_H\right)=20$ (cm$^{-2}$).
This is strong evidence that the two absorbers are one and the same. The X-ray
data also reveals a high-ionization phase at $\log\left(U_H\right)=1$, which makes
the separation in $N_H$ between the two phases similar to the one we see in
HE0238-1904. In NGC~3783 \citet{Gabel05b} find a single ionization solution for each of four UV 
outflow components spreading over $-1.6<\log\left(U_H\right)<-0.5$
and 3 ionization phases for the warm absorber with $-0.5<\log\left(U_H\right)<1.3$. 

The highest ionization phase detected in NGC~3783 ($\log\left(U_H\right)=1.3$) is inferred 
from troughs of the highly ionized species 
\sixiv\ (IP=2670 eV) and \sxvi\ (IP=3500 eV). 
\citet{Netzer03} find this phase to have $\log\left(U_H\right)$ 0.8 dex higher and $\log\left(U_H\right)$ 0.3 dex higher
than the medium ionization phase that can be detected using \neviii\ 
and \mgx\ measurements.  If a similar situation occurs in HE0238-1904 
the total mass flux and kinetic luminosity of the outflowing components (A and B) will increase
by a factor of 3 and the full width of the outflow will increase by 1.1 dex, and the filling factor of the low-ionization phase will drop accordingly.

Finally, we note the curious coincidence where for both outflow components in HE0238-1904, the two ionization phases fall 
along the \hi\ contour on the $N_H:U_H$ plot (see Figure \ref{fig:PhasePlotA}), 
or equivalently, have roughly a constant ratio of $N_H/U_H$.
This situation arises mainly due to the small measured $N_{ion}$ for \oiv\ (and \niv),
 contrasted with the much larger values (20-100 times, see table \ref{tabmeas}) for 
\neviii\ and \mgx.  A priori, our data is sensitive to a few percent of the measured  \neviii\ $N_{ion}$.  
Therefore, in principle such smaller measurements would have lowered the $N_H$ of the phase 2 by a similar amount, 
and thus the inferred situation is not a selection effect.

\section{DISCUSSION}\label{sec:discussion}

Significant AGN feedback processes typically require a mechanical energy input of roughly
0.5--5\% of the Eddington luminosity of the quasar \citep[][respectively]{Hopkins10,Scannapieco04}.
HE0238--1904, being among the brightest at its redshift band radiates close to its Eddington limit (i.e. $L_{Bol} \simeq L_{edd}$).
Therefore, with $\dot{E}_k\gtorder 0.7\%  L_{Bol}$, its outflow can contribute significantly to the theoretically invoked
AGN feedback processes.

Prior to the work described here, our group have shown that $\dot{E}_k$ of outflows in low-ionization 
BALQSO outflows are significant in the context of AGN feedback \citep{Moe09,Dunn10a}; as well as for the ubiquitous 
high ionization BALQSO \citep{Borguet13}. This investigation gives the first reliable estimates of $R$, $\dot{M}$ and $\dot{E}_k$
for the very high-ionization, high-luminosity quasar outflows. Cumulatively, these investigations suggest that UV absorption 
outflows observed in the rest-frame UV of luminous quasars, have enough kinetic luminosity to drive AGN feedback processes.
In Table \ref{tabmflo} we give the data for 3 representative types of quasar outflow: trough B in  HE0238-1904 from this analysis; 
the high ionization outflow in SDSS J1106+1939 from \citet{Borguet13}; and the low-ionization outflow in SDSS J0838+2955 from \citet{Moe09}. 

Other works have shown high mass flux ($\dot{M}$) for molecular outflows in luminous quasars. For example, \citet{Maiolino12}
shows such a large scale ($\sim$16 kpc) outflow at the very high redshift (z=6.4189) quasar J1148+5251. The outflow is reported to have 
$\dot{M}>3500 M_{\odot}$ yr$^{-1}$, but in order to have a fair comparison with $\dot{M}$ values we give  in Table \ref{tabmflo}, we need to use our 
$\dot{M}$ prescription (see eq.~\ref{energetics}) on the mass estimate from \citet{Maiolino12} (their eq.~2) which yields a third of the 
original $\dot{M}$ estimate (i.e., $\sim1200  M_{\odot}$ yr$^{-1}$).

The distances found for the two outflows we report here
are a few kiloparsecs from the central source. These are similar to the distances inferred for outflows in which
the density diagnostic is obtained from the study of excited troughs of singly ionized species as \feii\ or \siII\
\citep[e.g.][]{Korista08,Moe09,Dunn10a}, but they are 4+ orders of magnitude farther away than the assumed acceleration region
(0.03--0.1 pc) of line driven winds in quasars \citep[e.g.][]{Murray95,Proga00}. This result is consistent with
almost all the distances reported for AGN outflows in the literature. Our current investigation, expands this
claim to very high ionization outflows as well.  We conclude that most AGN outflows are observed very
far from their initial assumed acceleration region.

\begin{deluxetable}{lccrrrrrrr}
%\tabletypesize{\tiny}
\rotate
\tablecaption{{\sc Physical properties of the \siv\ quasar outflows.}}
\tablewidth{0pt}
\tablehead{
\colhead{Object}&\colhead{$\log(L_{Bol})$}&\colhead{$v$ }&\colhead{$\log(U_H)$ }&\colhead{$\log(N_H)$}&\colhead{$\log(n_e)$}&\colhead{$R$ }&\colhead{$\dot{M}$}&\colhead{$\log(\dot{E}_k)$}&\colhead{$\dot{E}_k/L_{Bol}$}\\
\colhead{}&\colhead{(ergs s$^{-1}$)}&\colhead{(km s$^{-1}$)}&\colhead{}&\colhead{($\mathrm{cm}^{-2}$)}&\colhead{($\mathrm{cm}^{-3}$)}&\colhead{(kpc)}&\colhead{(M$_{\odot}$ yr$^{-1}$)}&\colhead{(ergs s$^{-1}$)}&\colhead{(\%)}
}

\startdata

HE0238--1904 B $4Z_\odot$&$47.2$&$-5000$&$-2.5^{+0.1}_{-0.4}$&$20^{+0.1}_{-0.1}$&$3.75^{+0.10}_{-0.10}$&$3.4^{+2}_{-0.49}$&$31^{+20}_{-9}$&$45^{+0.2}_{-0.2}$&$0.7^{+0.5}_{-0.2}$\\

SDSS J1106+1939 $4Z_\odot$ & 47.2 & $-8250$ &
-0.5$^{+0.3}_{-0.2}$ & 22.1$^{+0.3}_{-0.1}$ &
4.1$^{+0.02}_{-0.37}$  & 0.32$^{+0.20}_{-0.10}$    &
390$^{+300}_{-10}$  & 46.0$^{+0.3}_{-0.1}$  &
5$^{+4}_{-0.3}$ \\

SDSS J0838+2955 & 47.5 & $-5000$ & $-2.0^{+0.2}_{-0.2}$&
20.8$^{+0.3}_{-0.3}$ & 3.8 & 3.3$^{+1.5}_{-1.0}$ & 300$^{+210
}_{-120}$ & 45.4$^{+0.2 }_{-0.2}$ & 0.8$^{+0.5
}_{-0.3}$ \\

\enddata

\label{tabmflo}
%c) Computed using $\Omega =0.2$ (see text for discussion).  ~\mathrm{c}
\end{deluxetable}

\clearpage

\section{SUMMARY}

In this paper we first explain the immense diagnostic power 
of spectra at rest-frame 500\AA--1000\AA, for quasar outflows science (see section 2 and Fig.~1), which includes: \\
a) Sensitivity to the warm absorber phase of the outflow: from observing  \neviii, \mgx\ and \sixii\ troughs. \\
b) Determining the distance from the AGN for the majority of high-ioinization outflows, 
by observing resonance and metastable troughs from such species (e.g., \oiv, \ov, \nev, \nevi) \\
c) Separating abundances and photoionization effects: by measuring troughs of several ions from the same element (e.g., \oii, \oiii, \oiv, \ov). 

The Cosmic Origins Spectrograph (COS) onboard HST allows us to realize this powerful diagnostic potential.
We can now obtain high enough quality data on bright medium-redshift quasars ($1.5\gtorder z \gtorder 0.5$), 
yielding detailed and reliable analysis of quasar-outflows spectra in this rest-wavelength regime. We demonstrate this potential by
 analyzing HST/COS (and FUSE) data of the outflow seen in quasar HE~0238--1904.
In the available data at rest-frame 600\AA--1100\AA\ we identify and measure a small subset of 
these diagnostic troughs, which yielded the following important results: \\
\noindent {\bf 1. Distance of the outflow from the AGN}: measuring the column densities ($N_{ion}$) of \oiv/\oiv*~$\lambda\lambda787.711,790.199$, 
yielded a distance of $\sim$ 3000 pc.\ for the outflow from the AGN. 
This is only the second time in the literature where a distance was determined for the majority of high ionization
quasar outflows (the first determination was also done by our group, see \citen{Borguet13} and table \ref{tabmflo} here). \\
   \noindent  {\bf 2. Two co-spatial ionization
phases:}  We measure $N_{ion}$ of troughs from the very high ionization species \neviii\ and \mgx, 
as well as from  the lower ionization species \oiv, \ov, \ovi, \niv, and \hi. To reproduce all of these $N_{ion}$,
photoionization modelling require two ionization phases in the outflow, 
with ionization parameters separated by roughly a factor of 100 between the two phases.
The low ionization phase is inferred to have a volume filling factor of $10^{-5}-10^{-6}$. \\
   \noindent  {\bf 3. Mass flux and kinetic luminosity of the outflow:} the two findings above, allow us to determine robust values 
for the : 40 \Msun\ yr$^{-1}$
and $10^{45}$ ergs s$^{-1}$, respectively, where the latter is roughly
equal to 1\% of the bolometric luminosity. 
Such a large kinetic luminosity and mass flux measured in a typical high
ionization wind, suggest that quasar outflows are a major contributor to AGN
feedback mechanisms.

\section*{ACKNOWLEDGMENTS}
            
 We acknowledge support from NASA STScI grants GO 11686 and GO
12022 as well as NSF grant AST 0837880. We thank Sowgat Muzahid for helpful suggestions on this manuscript.

\bibliographystyle{apj}
\bibliography{astro}{}

\begin{thebibliography}{80}
\expandafter\ifx\csname natexlab\endcsname\relax\def\natexlab#1{#1}\fi

\bibitem[{{Aoki} {et~al.}(2011){Aoki}, {Oyabu}, {Dunn}, {Arav}, {Edmonds},
  {Korista}, {Matsuhara}, \& {Toba}}]{Aoki11}
{Aoki}, K., {Oyabu}, S., {Dunn}, J.~P., {Arav}, N., {Edmonds}, D., {Korista},
  K.~T., {Matsuhara}, H., \& {Toba}, Y. 2011, \pasj, 63, 457

\bibitem[{{Arav}(1997)}]{Arav97}
{Arav}, N. 1997, in Astronomical Society of the Pacific Conference Series, Vol.
  128, Mass Ejection from Active Galactic Nuclei, ed. N.~{Arav}, I.~{Shlosman},
  \& R.~J. {Weymann}, 208

\bibitem[{{Arav} {et~al.}(1999{\natexlab{a}}){Arav}, {Becker},
  {Laurent-Muehleisen}, {Gregg}, {White}, {Brotherton}, \& {de Kool}}]{Arav99b}
{Arav}, N., {Becker}, R.~H., {Laurent-Muehleisen}, S.~A., {Gregg}, M.~D.,
  {White}, R.~L., {Brotherton}, M.~S., \& {de Kool}, M. 1999{\natexlab{a}},
  \apj, 524, 566

\bibitem[{{Arav} {et~al.}(2001{\natexlab{a}}){Arav}, {Brotherton}, {Becker},
  {Gregg}, {White}, {Price}, \& {Hack}}]{Arav01b}
{Arav}, N., {Brotherton}, M.~S., {Becker}, R.~H., {Gregg}, M.~D., {White},
  R.~L., {Price}, T., \& {Hack}, W. 2001{\natexlab{a}}, \apj, 546, 140

\bibitem[{{Arav} {et~al.}(2005){Arav}, {Kaastra}, {Kriss}, {Korista}, {Gabel},
  \& {Proga}}]{Arav05}
{Arav}, N., {Kaastra}, J., {Kriss}, G.~A., {Korista}, K.~T., {Gabel}, J., \&
  {Proga}, D. 2005, \apj, 620, 665

\bibitem[{{Arav} {et~al.}(2003){Arav}, {Kaastra}, {Steenbrugge}, {Brinkman},
  {Edelson}, {Korista}, \& {de Kool}}]{Arav03}
{Arav}, N., {Kaastra}, J., {Steenbrugge}, K., {Brinkman}, B., {Edelson}, R.,
  {Korista}, K.~T., \& {de Kool}, M. 2003, \apj, 590, 174

\bibitem[{{Arav} {et~al.}(2002){Arav}, {Korista}, \& {de Kool}}]{Arav02}
{Arav}, N., {Korista}, K.~T., \& {de Kool}, M. 2002, \apj, 566, 699

\bibitem[{{Arav} {et~al.}(1999{\natexlab{b}}){Arav}, {Korista}, {de Kool},
  {Junkkarinen}, \& {Begelman}}]{Arav99a}
{Arav}, N., {Korista}, K.~T., {de Kool}, M., {Junkkarinen}, V.~T., \&
  {Begelman}, M.~C. 1999{\natexlab{b}}, \apj, 516, 27

\bibitem[{{Arav} {et~al.}(2008){Arav}, {Moe}, {Costantini}, {Korista}, {Benn},
  \& {Ellison}}]{Arav08}
{Arav}, N., {Moe}, M., {Costantini}, E., {Korista}, K.~T., {Benn}, C., \&
  {Ellison}, S. 2008, \apj, 681, 954

\bibitem[{{Arav} {et~al.}(2001{\natexlab{b}}){Arav}, {de Kool}, {Korista},
  {Crenshaw}, {van Breugel}, {Brotherton}, {Green}, {Pettini}, {Wills}, {de
  Vries}, {Becker}, {Brandt}, {Green}, {Junkkarinen}, {Koratkar}, {Laor},
  {Laurent-Muehleisen}, {Mathur}, \& {Murray}}]{Arav01}
{Arav}, N., {et~al.} 2001{\natexlab{b}}, \apj, 561, 118

\bibitem[{{Arav} {et~al.}(2007){Arav}, {Gabel}, {Korista}, {Kaastra}, {Kriss},
  {Behar}, {Costantini}, {Gaskell}, {Laor}, {Kodituwakku}, {Proga}, {Sako},
  {Scott}, \& {Steenbrugge}}]{Arav07}
---. 2007, \apj, 658, 829

\bibitem[{{Arav} {et~al.}(2012){Arav}, {Edmonds}, {Borguet}, {Kriss},
  {Kaastra}, {Behar}, {Bianchi}, {Cappi}, {Costantini}, {Detmers}, {Ebrero},
  {Mehdipour}, {Paltani}, {Petrucci}, {Pinto}, {Ponti}, {Steenbrugge}, \& {de
  Vries}}]{Arav12}
---. 2012, \aap, 544, A33

\bibitem[{{Barlow} {et~al.}(1997){Barlow}, {Hamann}, \& {Sargent}}]{Barlow97a}
{Barlow}, T.~A., {Hamann}, F., \& {Sargent}, W.~L.~W. 1997, in Astronomical
  Society of the Pacific Conference Series, Vol. 128, Mass Ejection from Active
  Galactic Nuclei, ed. {N.~Arav, I.~Shlosman, \& R.~J.~Weymann}, 13--+

\bibitem[{{Bautista} {et~al.}(2010){Bautista}, {Dunn}, {Arav}, {Korista},
  {Moe}, \& {Benn}}]{Bautista10}
{Bautista}, M.~A., {Dunn}, J.~P., {Arav}, N., {Korista}, K.~T., {Moe}, M., \&
  {Benn}, C. 2010, \apj, 713, 25

\bibitem[{{Borguet} {et~al.}(2013){Borguet}, {Arav}, {Edmonds}, {Chamberlain},
  \& {Benn}}]{Borguet13}
{Borguet}, B.~C.~J., {Arav}, N., {Edmonds}, D., {Chamberlain}, C., \& {Benn},
  C. 2013, \apj, 762, 49

\bibitem[{{Borguet} {et~al.}(2012{\natexlab{a}}){Borguet}, {Edmonds}, {Arav},
  {Benn}, \& {Chamberlain}}]{Borguet12b}
{Borguet}, B.~C.~J., {Edmonds}, D., {Arav}, N., {Benn}, C., \& {Chamberlain},
  C. 2012{\natexlab{a}}, \apj, 758, 69

\bibitem[{{Borguet} {et~al.}(2012{\natexlab{b}}){Borguet}, {Edmonds}, {Arav},
  {Dunn}, \& {Kriss}}]{Borguet12a}
{Borguet}, B.~C.~J., {Edmonds}, D., {Arav}, N., {Dunn}, J., \& {Kriss}, G.~A.
  2012{\natexlab{b}}, \apj, 751, 107

\bibitem[{{Cattaneo} {et~al.}(2009){Cattaneo}, {Faber}, {Binney}, {Dekel},
  {Kormendy}, {Mushotzky}, {Babul}, {Best}, {Br{\"u}ggen}, {Fabian}, {Frenk},
  {Khalatyan}, {Netzer}, {Mahdavi}, {Silk}, {Steinmetz}, \&
  {Wisotzki}}]{Cattaneo09}
{Cattaneo}, A., {et~al.} 2009, \nat, 460, 213

\bibitem[{{Churchill} {et~al.}(1999){Churchill}, {Mellon}, {Charlton},
  {Jannuzi}, {Kirhakos}, {Steidel}, \& {Schneider}}]{Churchill99}
{Churchill}, C.~W., {Mellon}, R.~R., {Charlton}, J.~C., {Jannuzi}, B.~T.,
  {Kirhakos}, S., {Steidel}, C.~C., \& {Schneider}, D.~P. 1999, \apjl, 519, L43

\bibitem[{{Ciotti} {et~al.}(2009){Ciotti}, {Ostriker}, \& {Proga}}]{Ciotti09}
{Ciotti}, L., {Ostriker}, J.~P., \& {Proga}, D. 2009, \apj, 699, 89

\bibitem[{{Ciotti} {et~al.}(2010){Ciotti}, {Ostriker}, \& {Proga}}]{Ciotti10}
---. 2010, \apj, 717, 708

\bibitem[{{Costantini} {et~al.}(2007){Costantini}, {Kaastra}, {Arav}, {Kriss},
  {Steenbrugge}, {Gabel}, {Verbunt}, {Behar}, {Gaskell}, {Korista}, {Proga},
  {Quijano}, {Scott}, {Klimek}, \& {Hedrick}}]{Costantini07}
{Costantini}, E., {et~al.} 2007, \aap, 461, 121

\bibitem[{{Dai} {et~al.}(2008){Dai}, {Shankar}, \& {Sivakoff}}]{Dai08}
{Dai}, X., {Shankar}, F., \& {Sivakoff}, G.~R. 2008, \apj, 672, 108

\bibitem[{{Dai} {et~al.}(2012){Dai}, {Shankar}, \& {Sivakoff}}]{Dai12}
---. 2012, \apj, 757, 180

\bibitem[{{Danforth} {et~al.}(2010){Danforth}, {Keeney}, {Stocke}, {Shull}, \&
  {Yao}}]{Danforth10}
{Danforth}, C.~W., {Keeney}, B.~A., {Stocke}, J.~T., {Shull}, J.~M., \& {Yao},
  Y. 2010, \apj, 720, 976

\bibitem[{{de Kool} {et~al.}(2001){de Kool}, {Arav}, {Becker}, {Gregg},
  {White}, {Laurent-Muehleisen}, {Price}, \& {Korista}}]{deKool01}
{de Kool}, M., {Arav}, N., {Becker}, R.~H., {Gregg}, M.~D., {White}, R.~L.,
  {Laurent-Muehleisen}, S.~A., {Price}, T., \& {Korista}, K.~T. 2001, \apj,
  548, 609

\bibitem[{{de Kool} {et~al.}(2002{\natexlab{a}}){de Kool}, {Becker}, {Gregg},
  {White}, \& {Arav}}]{deKool02b}
{de Kool}, M., {Becker}, R.~H., {Gregg}, M.~D., {White}, R.~L., \& {Arav}, N.
  2002{\natexlab{a}}, \apj, 567, 58

\bibitem[{{de Kool} \& {Begelman}(1995)}]{dekool95}
{de Kool}, M., \& {Begelman}, M.~C. 1995, \apj, 455, 448

\bibitem[{{de Kool} {et~al.}(2002{\natexlab{b}}){de Kool}, {Korista}, \&
  {Arav}}]{deKool02}
{de Kool}, M., {Korista}, K.~T., \& {Arav}, N. 2002{\natexlab{b}}, \apj, 580,
  54

\bibitem[{{Di Matteo} {et~al.}(2005){Di Matteo}, {Springel}, \&
  {Hernquist}}]{dimatteo05}
{Di Matteo}, T., {Springel}, V., \& {Hernquist}, L. 2005, \nat, 433, 604

\bibitem[{{Dixon} {et~al.}(2007){Dixon}, {Sahnow}, {Barrett}, {Civeit},
  {Dupuis}, {Fullerton}, {Godard}, {Hsu}, {Kaiser}, {Kruk}, {Lacour},
  {Lindler}, {Massa}, {Robinson}, {Romelfanger}, \& {Sonnentrucker}}]{Dixon07}
{Dixon}, W.~V., {et~al.} 2007, \pasp, 119, 527

\bibitem[{{Dunn} {et~al.}(2012){Dunn}, {Arav}, {Aoki}, {Wilkins}, {Laughlin},
  {Edmonds}, \& {Bautista}}]{Dunn12}
{Dunn}, J.~P., {Arav}, N., {Aoki}, K., {Wilkins}, A., {Laughlin}, C.,
  {Edmonds}, D., \& {Bautista}, M. 2012, \apj, 750, 143

\bibitem[{{Dunn} {et~al.}(2010{\natexlab{a}}){Dunn}, {Crenshaw}, {Kraemer}, \&
  {Trippe}}]{Dunn10b}
{Dunn}, J.~P., {Crenshaw}, D.~M., {Kraemer}, S.~B., \& {Trippe}, M.~L.
  2010{\natexlab{a}}, \apj, 713, 900

\bibitem[{{Dunn} {et~al.}(2010{\natexlab{b}}){Dunn}, {Bautista}, {Arav}, {Moe},
  {Korista}, {Costantini}, {Benn}, {Ellison}, \& {Edmonds}}]{Dunn10a}
{Dunn}, J.~P., {et~al.} 2010{\natexlab{b}}, \apj, 709, 611

\bibitem[{{Edmonds} {et~al.}(2011){Edmonds}, {Borguet}, {Arav}, {Dunn},
  {Penton}, {Kriss}, {Korista}, {Costantini}, {Steenbrugge},
  {Gonzalez-Serrano}, {Aoki}, {Bautista}, {Behar}, {Benn}, {Crenshaw},
  {Everett}, {Gabel}, {Kaastra}, {Moe}, \& {Scott}}]{Edmonds11}
{Edmonds}, D., {et~al.} 2011, \apj, 739, 7

\bibitem[{{Elvis}(2006)}]{Elvis06}
{Elvis}, M. 2006, \memsai, 77, 573

\bibitem[{{Faucher-Gigu{\`e}re} {et~al.}(2012){Faucher-Gigu{\`e}re},
  {Quataert}, \& {Murray}}]{Faucher-Giguere12}
{Faucher-Gigu{\`e}re}, C.-A., {Quataert}, E., \& {Murray}, N. 2012, \mnras,
  420, 1347

\bibitem[{{Ferland} {et~al.}(1998){Ferland}, {Korista}, {Verner}, {Ferguson},
  {Kingdon}, \& {Verner}}]{Ferland98}
{Ferland}, G.~J., {Korista}, K.~T., {Verner}, D.~A., {Ferguson}, J.~W.,
  {Kingdon}, J.~B., \& {Verner}, E.~M. 1998, \pasp, 110, 761

\bibitem[{{Gabel} {et~al.}(2006){Gabel}, {Arav}, \& {Kim}}]{Gabel06}
{Gabel}, J.~R., {Arav}, N., \& {Kim}, T. 2006, \apj, 646, 742

\bibitem[{{Gabel} {et~al.}(2003){Gabel}, {Crenshaw}, {Kraemer}, {Brandt},
  {George}, {Hamann}, {Kaiser}, {Kaspi}, {Kriss}, {Mathur}, {Mushotzky},
  {Nandra}, {Netzer}, {Peterson}, {Shields}, {Turner}, \& {Zheng}}]{Gabel03}
{Gabel}, J.~R., {et~al.} 2003, \apj, 583, 178

\bibitem[{{Gabel} {et~al.}(2005{\natexlab{a}}){Gabel}, {Kraemer}, {Crenshaw},
  {George}, {Brandt}, {Hamann}, {Kaiser}, {Kaspi}, {Kriss}, {Mathur}, {Nandra},
  {Netzer}, {Peterson}, {Shields}, {Turner}, \& {Zheng}}]{Gabel05b}
---. 2005{\natexlab{a}}, \apj, 631, 741

\bibitem[{{Gabel} {et~al.}(2005{\natexlab{b}}){Gabel}, {Arav}, {Kaastra},
  {Kriss}, {Behar}, {Costantini}, {Gaskell}, {Korista}, {Laor}, {Paerels},
  {Proga}, {Quijano}, {Sako}, {Scott}, \& {Steenbrugge}}]{Gabel05a}
---. 2005{\natexlab{b}}, \apj, 623, 85

\bibitem[{{Ganguly} \& {Brotherton}(2008)}]{Ganguly08}
{Ganguly}, R., \& {Brotherton}, M.~S. 2008, \apj, 672, 102

\bibitem[{{Ganguly} {et~al.}(1999){Ganguly}, {Eracleous}, {Charlton}, \&
  {Churchill}}]{Ganguly99}
{Ganguly}, R., {Eracleous}, M., {Charlton}, J.~C., \& {Churchill}, C.~W. 1999,
  \aj, 117, 2594

\bibitem[{{Ganguly} {et~al.}(2006){Ganguly}, {Sembach}, {Tripp}, {Savage}, \&
  {Wakker}}]{Ganguly06}
{Ganguly}, R., {Sembach}, K.~R., {Tripp}, T.~M., {Savage}, B.~D., \& {Wakker},
  B.~P. 2006, \apj, 645, 868

\bibitem[{{Germain} {et~al.}(2009){Germain}, {Barai}, \& {Martel}}]{Germain09}
{Germain}, J., {Barai}, P., \& {Martel}, H. 2009, \apj, 704, 1002

\bibitem[{{Hall} {et~al.}(2002){Hall}, {Anderson}, {Strauss}, {York},
  {Richards}, {Fan}, {Knapp}, {Schneider}, {Vanden Berk}, {Geballe}, {Bauer},
  {Becker}, {Davis}, {Rix}, {Nichol}, {Bahcall}, {Brinkmann}, {Brunner},
  {Connolly}, {Csabai}, {Doi}, {Fukugita}, {Gunn}, {Haiman}, {Harvanek},
  {Heckman}, {Hennessy}, {Inada}, {Ivezi{\'c}}, {Johnston}, {Kleinman},
  {Krolik}, {Krzesinski}, {Kunszt}, {Lamb}, {Long}, {Lupton}, {Miknaitis},
  {Munn}, {Narayanan}, {Neilsen}, {Newman}, {Nitta}, {Okamura}, {Pentericci},
  {Pier}, {Schlegel}, {Snedden}, {Szalay}, {Thakar}, {Tsvetanov}, {White}, \&
  {Zheng}}]{Hall02}
{Hall}, P.~B., {et~al.} 2002, \apjs, 141, 267

\bibitem[{{Hamann} {et~al.}(1995){Hamann}, {Barlow}, {Beaver}, {Burbidge},
  {Cohen}, {Junkkarinen}, \& {Lyons}}]{Hamann95}
{Hamann}, F., {Barlow}, T.~A., {Beaver}, E.~A., {Burbidge}, E.~M., {Cohen},
  R.~D., {Junkkarinen}, V., \& {Lyons}, R. 1995, \apj, 443, 606

\bibitem[{{Hamann} {et~al.}(1997){Hamann}, {Barlow}, {Junkkarinen}, \&
  {Burbidge}}]{Hamann97b}
{Hamann}, F., {Barlow}, T.~A., {Junkkarinen}, V., \& {Burbidge}, E.~M. 1997,
  \apj, 478, 80

\bibitem[{{Hamann} \& {Ferland}(1993)}]{Hamann93}
{Hamann}, F., \& {Ferland}, G. 1993, \apj, 418, 11

\bibitem[{{Hamann} {et~al.}(2001){Hamann}, {Barlow}, {Chaffee}, {Foltz}, \&
  {Weymann}}]{Hamann01}
{Hamann}, F.~W., {Barlow}, T.~A., {Chaffee}, F.~C., {Foltz}, C.~B., \&
  {Weymann}, R.~J. 2001, \apj, 550, 142

\bibitem[{{Hamann} {et~al.}(2000){Hamann}, {Netzer}, \& {Shields}}]{Hamann00}
{Hamann}, F.~W., {Netzer}, H., \& {Shields}, J.~C. 2000, \apj, 536, 101

\bibitem[{{Hewett} \& {Foltz}(2003)}]{Hewett03}
{Hewett}, P.~C., \& {Foltz}, C.~B. 2003, \aj, 125, 1784

\bibitem[{{Holczer} {et~al.}(2007){Holczer}, {Behar}, \& {Kaspi}}]{Holczer07}
{Holczer}, T., {Behar}, E., \& {Kaspi}, S. 2007, \apj, 663, 799

\bibitem[{{Hopkins} \& {Elvis}(2010)}]{Hopkins10}
{Hopkins}, P.~F., \& {Elvis}, M. 2010, \mnras, 401, 7

\bibitem[{{Hopkins} {et~al.}(2006){Hopkins}, {Hernquist}, {Cox}, {Di Matteo},
  {Robertson}, \& {Springel}}]{Hopkins06}
{Hopkins}, P.~F., {Hernquist}, L., {Cox}, T.~J., {Di Matteo}, T., {Robertson},
  B., \& {Springel}, V. 2006, \apjs, 163, 1

\bibitem[{{Hopkins} {et~al.}(2009){Hopkins}, {Murray}, \&
  {Thompson}}]{Hopkins09}
{Hopkins}, P.~F., {Murray}, N., \& {Thompson}, T.~A. 2009, \mnras, 398, 303

\bibitem[{{Knigge} {et~al.}(2008){Knigge}, {Scaringi}, {Goad}, \&
  {Cottis}}]{Knigge08}
{Knigge}, C., {Scaringi}, S., {Goad}, M.~R., \& {Cottis}, C.~E. 2008, \mnras,
  386, 1426

\bibitem[{{Korista} {et~al.}(2008){Korista}, {Bautista}, {Arav}, {Moe},
  {Costantini}, \& {Benn}}]{Korista08}
{Korista}, K.~T., {Bautista}, M.~A., {Arav}, N., {Moe}, M., {Costantini}, E.,
  \& {Benn}, C. 2008, \apj, 688, 108

\bibitem[{{Korista} {et~al.}(1992){Korista}, {Weymann}, {Morris}, {Kopko},
  {Turnshek}, {Hartig}, {Foltz}, {Burbidge}, \& {Junkkarinen}}]{Korista92}
{Korista}, K.~T., {et~al.} 1992, \apj, 401, 529

\bibitem[{{Levine} \& {Gnedin}(2005)}]{Levine05}
{Levine}, R., \& {Gnedin}, N.~Y. 2005, \apj, 632, 727

\bibitem[{{Maiolino} {et~al.}(2012){Maiolino}, {Gallerani}, {Neri}, {Cicone},
  {Ferrara}, {Genzel}, {Lutz}, {Sturm}, {Tacconi}, {Walter}, {Feruglio},
  {Fiore}, \& {Piconcelli}}]{Maiolino12}
{Maiolino}, R., {et~al.} 2012, \mnras, 425, L66

\bibitem[{{Mathews} \& {Ferland}(1987)}]{Mathews87}
{Mathews}, W.~G., \& {Ferland}, G.~J. 1987, \apj, 323, 456

\bibitem[{{Moe} {et~al.}(2009){Moe}, {Arav}, {Bautista}, \& {Korista}}]{Moe09}
{Moe}, M., {Arav}, N., {Bautista}, M.~A., \& {Korista}, K.~T. 2009, \apj, 706,
  525

\bibitem[{{Murray} {et~al.}(1995){Murray}, {Chiang}, {Grossman}, \&
  {Voit}}]{Murray95}
{Murray}, N., {Chiang}, J., {Grossman}, S.~A., \& {Voit}, G.~M. 1995, \apj,
  451, 498

\bibitem[{{Muzahid} {et~al.}(2013){Muzahid}, {Srianand}, {Arav}, {Savage}, \&
  {Narayanan}}]{Muzahid13}
{Muzahid}, S., {Srianand}, R., {Arav}, N., {Savage}, B.~D., \& {Narayanan}, A.
  2013, \mnras, 431, 2885

\bibitem[{{Muzahid} {et~al.}(2012){Muzahid}, {Srianand}, {Savage}, {Narayanan},
  {Mohan}, \& {Dewangan}}]{Muzahid12}
{Muzahid}, S., {Srianand}, R., {Savage}, B.~D., {Narayanan}, A., {Mohan}, V.,
  \& {Dewangan}, G.~C. 2012, \mnras, 424, L59

\bibitem[{{Netzer} {et~al.}(2003){Netzer}, {Kaspi}, {Behar}, {Brandt},
  {Chelouche}, {George}, {Crenshaw}, {Gabel}, {Hamann}, {Kraemer}, {Kriss},
  {Nandra}, {Peterson}, {Shields}, \& {Turner}}]{Netzer03}
{Netzer}, H., {et~al.} 2003, \apj, 599, 933

\bibitem[{{Osterman} {et~al.}(2010){Osterman}, {Green}, {Froning},
  {B{\'e}land}, {Burgh}, {France}, {Penton}, {Delker}, {Ebbets}, {Sahnow},
  {Bacinski}, {Kimble}, {Andrews}, {Wilkinson}, {McPhate}, {Siegmund}, {Ake},
  {Aloisi}, {Biagetti}, {Diaz}, {Dixon}, {Friedman}, {Ghavamian}, {Goudfrooij},
  {Hartig}, {Keyes}, {Lennon}, {Massa}, {Niemi}, {Oliveira}, {Osten},
  {Proffitt}, {Smith}, \& {Soderblom}}]{Osterman10}
{Osterman}, S., {et~al.} 2010, arXiv:1012.5827 [astro-ph.IM]

\bibitem[{{Ostriker} {et~al.}(2010){Ostriker}, {Choi}, {Ciotti}, {Novak}, \&
  {Proga}}]{Ostriker10}
{Ostriker}, J.~P., {Choi}, E., {Ciotti}, L., {Novak}, G.~S., \& {Proga}, D.
  2010, \apj, 722, 642

\bibitem[{{Petitjean} \& {Srianand}(1999)}]{Petitjean99}
{Petitjean}, P., \& {Srianand}, R. 1999, \aap, 345, 73

\bibitem[{{Proga} {et~al.}(2000){Proga}, {Stone}, \& {Kallman}}]{Proga00}
{Proga}, D., {Stone}, J.~M., \& {Kallman}, T.~R. 2000, \apj, 543, 686

\bibitem[{{Scannapieco} \& {Oh}(2004)}]{Scannapieco04}
{Scannapieco}, E., \& {Oh}, S.~P. 2004, \apj, 608, 62

\bibitem[{{Schlegel} {et~al.}(1998){Schlegel}, {Finkbeiner}, \&
  {Davis}}]{Schlegel98}
{Schlegel}, D.~J., {Finkbeiner}, D.~P., \& {Davis}, M. 1998, \apj, 500, 525

\bibitem[{{Scott} {et~al.}(2004){Scott}, {Kriss}, {Lee}, {Arav}, {Ogle},
  {Roraback}, {Weaver}, {Alexander}, {Brotherton}, {Green}, {Hutchings},
  {Kaiser}, {Marshall}, {Oegerle}, \& {Zheng}}]{Scott04}
{Scott}, J.~E., {et~al.} 2004, \apjs, 152, 1

\bibitem[{{Silk} \& {Rees}(1998)}]{Silk98}
{Silk}, J., \& {Rees}, M.~J. 1998, \aap, 331, L1

\bibitem[{{Steenbrugge} {et~al.}(2005){Steenbrugge}, {Kaastra}, {Crenshaw},
  {Kraemer}, {Arav}, {George}, {Liedahl}, {van der Meer}, {Paerels}, {Turner},
  \& {Yaqoob}}]{Steenbrugge05}
{Steenbrugge}, K.~C., {et~al.} 2005, \aap, 434, 569

\bibitem[{{Telfer} {et~al.}(1998){Telfer}, {Kriss}, {Zheng}, {Davidsen}, \&
  {Green}}]{Telfer98}
{Telfer}, R.~C., {Kriss}, G.~A., {Zheng}, W., {Davidsen}, A.~F., \& {Green},
  R.~F. 1998, \apj, 509, 132

\bibitem[{{Telfer} {et~al.}(2002){Telfer}, {Zheng}, {Kriss}, \&
  {Davidsen}}]{Telfer02}
{Telfer}, R.~C., {Zheng}, W., {Kriss}, G.~A., \& {Davidsen}, A.~F. 2002, \apj,
  565, 773

\bibitem[{{Wampler} {et~al.}(1995){Wampler}, {Chugai}, \&
  {Petitjean}}]{Wampler95}
{Wampler}, E.~J., {Chugai}, N.~N., \& {Petitjean}, P. 1995, \apj, 443, 586

\end{thebibliography}
%\bibliography{astro}
            
\end{document}